\documentclass[sigconf, nonacm]{acmart}

\usepackage{amsmath}
\usepackage{graphicx}
\usepackage{balance}  %
\usepackage{adjustbox}
\usepackage{footmisc}
\usepackage{soul}
\usepackage{hyperref}
\usepackage{enumitem}
\usepackage{multirow}
\usepackage{listings}
\usepackage[noend]{algpseudocode}
\usepackage[ruled,vlined]{algorithm2e}
\usepackage{cleveref}
\usepackage{array}
\usepackage{tabularx}
\usepackage{booktabs} %
\usepackage{xspace}
\usepackage{xcolor}
\usepackage{microtype}
\usepackage{subcaption}
\usepackage{balance}
\usepackage{mathtools}
\usepackage{scalerel}
\usepackage{adjustbox}
\usepackage{float}
\usepackage{appendix}

\definecolor{darkgreen}{rgb}{0.15,0.55,0.15}
\definecolor{darkblue}{rgb}{0.1,0.1,0.5}
\definecolor{orange}{rgb}{1, 0.64, 0}
\definecolor{blue}{rgb}{0.01,0.40,.8}
\definecolor{darkgreen}{rgb}{0.15,0.55,0.15}
\definecolor{mred}{rgb}{.80,.12,.30}
\definecolor{grey}{rgb}{0.5,0.5,0.5}
\definecolor{Purple}{rgb}{.75,0,.85}
\definecolor{light-gray}{gray}{0.95}
\definecolor{mid-gray}{gray}{0.85}
\definecolor{darkred}{rgb}{0.7,0.25,0.25}
\definecolor{rose}{rgb}{1.0, 0.01, 0.24}
\definecolor{lightyellow}{HTML}{FFD700}

\newcommand{\gray}[1]{\textcolor{grey}{#1}}

\newcommand{\highlight}[1]{{\setlength{\fboxsep}{1pt}\colorbox{yellow}{\textbf{\texttt{#1}}}}}
\newcommand{\ghighlight}[1]{{\setlength{\fboxsep}{1pt}\colorbox{green}{\textbf{\texttt{#1}}}}}

\newcommand{\eat}[1]{}

\newcommand{\ititle}[1]{\vspace{2pt}\noindent\emph{#1}}
\newcommand{\stitle}[1]{\smallskip\noindent\textbf{#1}}

\newlength{\listingindent}                %
\usepackage[LGRgreek]{mathastext}

\newenvironment{myitemize}{\vspace{-0.1in}\begin{list}{\gray{$\bullet$}}{\renewcommand{\leftmargin}{0.15in}}}{\end{list}\vspace{-.1in}}

\newtheorem{example}{Example}

\newtheorem{problem}{Problem}

\newcommand{\sysold}{\textsc{PI$_1$}\xspace}

\newcommand{\sys}{\textsc{PI$_2$}\xspace}
\newcommand{\difftree}{\textsc{Difftree}\xspace}
\newcommand{\difftrees}{{\difftree}s\xspace}

\begin{document}

\title{\sys: End-to-end Interactive Visualization Interface \\Generation from Queries}

\subtitle{Technical Report\thanks{\cite{SIGMODPI, demo} are the published versions in SIGMOD 2022.}}

\author{Yiru Chen}
\affiliation{%
  \institution{Columbia University}
}
\email{yiru.chen@columbia.edu}

\author{Eugene Wu}
\affiliation{%
  \institution{DSI, Columbia University}
}
\email{ewu@cs.columbia.edu}

\begin{abstract}
Interactive visual analysis interfaces are critical in nearly every data
task. Yet creating new interfaces is deeply challenging, as it requires   
the developer to understand the queries needed to express the desired analysis task,
design the appropriate interface to express those queries for the task,
and implement the interface using a combination of visualization, browser, server, and database technologies.
Although prior work generates a set of interactive widgets that can express an input query log,
this paper presents \sys, the first system to generate fully functional visual analysis interfaces
from an example sequence of analysis queries.
\sys analyzes queries syntactically and represents a set of queries using a novel \difftree structure that encodes systematic variations between query abstract syntax trees.
\sys then maps each \difftree to a visualization that renders its results, the variations in each \difftree to interactions,
and generates a good layout for the interface.
We show that \sys can express data-oriented interactions in existing visualization
interaction taxonomies, can reproduce or improve several real-world visual analysis interfaces,
generates interfaces in $2 - 19$s (median 6s), and scales linearly with the number of queries.

\end{abstract}

\maketitle

\section{Introduction}\label{s:intro}

Interactive visual analysis interfaces (or simply {\it interfaces}) are 
critical in nearly every stage of data management, including data cleaning~\cite{wu2013scorpion}, wrangling~\cite{Kandel2011WranglerIV}, modeling~\cite{facets}, exploration~\cite{Murray2013TableauYD, TSExplain}, and communication~\cite{icheck,fivethirtyeight}.  
Interfaces empower the user to easily express relevant analysis queries using interactive controls that hide the underlying query complexity.
A prominent example is VizQL~\cite{Stolte2008PolarisAS} (commercialized as Tableau), which was carefully designed for analyses based on OLAP cube queries.  
However, analyses are not restricted to OLAP queries and can be arbitrarily complex.  
Thus, translating a custom analysis query workload into a fully functional interface remains deeply challenging.

\begin{figure}
  \centering
  \includegraphics[width=.9\columnwidth]{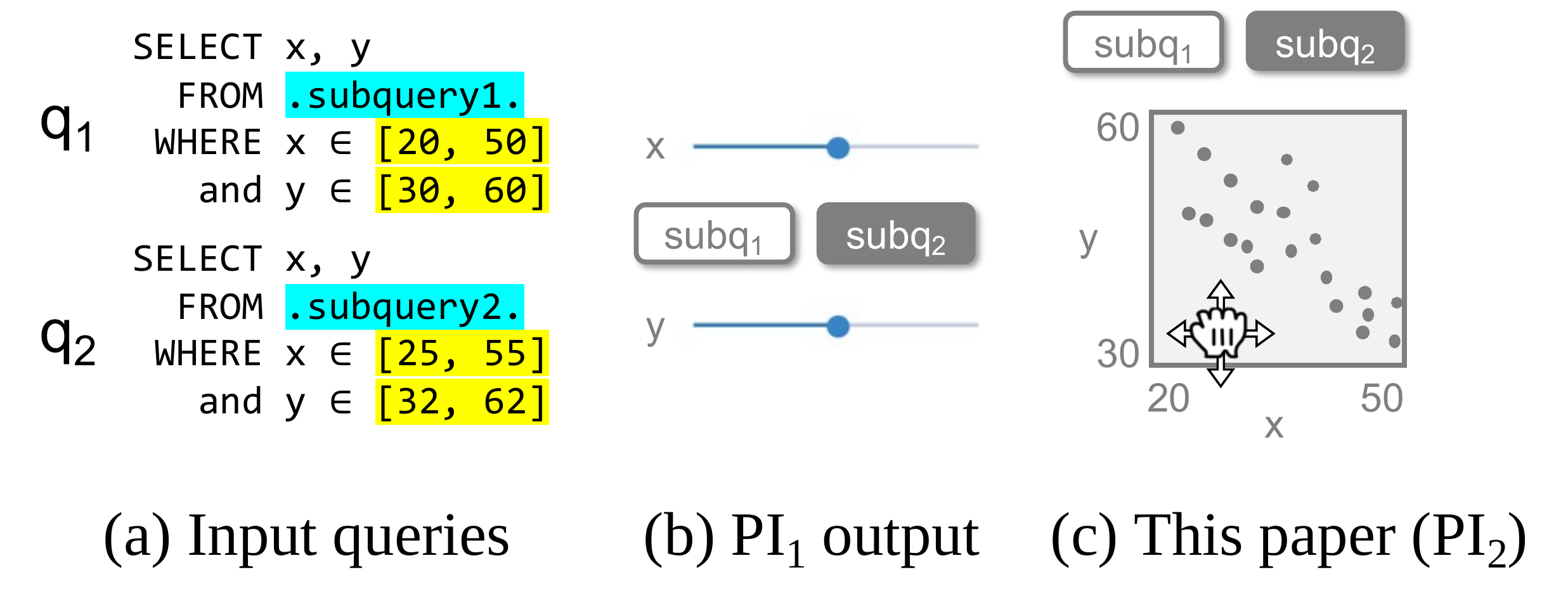}
  \vspace{-.1in}
  \caption{Comparison between prior and this work.  (b) Prior work generates 
    an unordered set of interaction widgets that express the two queries.
    (c) This paper presents a novel model that accounts for
      widgets, layouts, and interactive visualizations, and 
    generate fully interactive interfaces.}
  \label{f:intro}
\end{figure}

\begin{example}
  The two queries in \Cref{f:intro}(a) differ in two ways: the from clause chooses from two subqueries, and the two predicate ranges may change.  
  Designing a good interface requires numerous nuanced decisions.
  Should the queries be rendered together or separately?
  As scatterplots (supports panning along the y-axis), bar charts (which do not), or another chart type?
  How should the user choose the subquery?
  Should textboxes, sliders, range sliders, or panning specify the range predicates?
  Should the layout be horizontal or vertical?
  What if the screen is wide?  Or narrow?  
  \Cref{f:intro}(c) depicts a sensible interface that expresses these queries.
  It visualizes the query results in a scatter plot.  The user can drag the plot to change the
  x and y predicate ranges, and click on the buttons to select the desired subquery.  

  Once a design is established, the developer still needs to 
  ensure that interactions appropriately transform the underlying queries; 
  implement the interface using a mix of browser, visualization, server, and database technologies;
  debug her implementation; and finally deploy the result.  
  Only then, can she finally use or share the interface.
\end{example}

The example highlights how intimidating and laborious it is to navigate design decisions and build interfaces using a multitude of technologies.  
To this end, dashboard creators (e.g., Metabase~\cite{metabase}, Retool~\cite{retool}) and exploration tools (e.g., Tableau~\cite{Murray2013TableauYD}) help author SQL-based analysis dashboards,
however they limit the scope of analyses and queries in order to keep their own interfaces simple.  
For instance, Metabase is restricted to parameterized queries~\cite{metabasedoc}, and Tableau to OLAP cube queries~\cite{tableaucube}.    
Analyses that go beyond these restrictions will require manual implementation.  

To simplify interface implementation, numerous programming libraries have been created to recommend~\cite{wang2021falx,Zong2019Lyra2D} or create~\cite{Satyanarayan2017VegaLiteAG,Bostock2011DDD} visualizations, manipulate the DOM~\cite{jquery}, manage a web server~\cite{flask}, and construct SQL queries~\cite{knexjs}.
Similarly, tools such as AirTable~\cite{airtable}, Figma~\cite{figma}, and Plasmic~\cite{plasmic} focus on rapid interface design.  
Ultimately, each only solves one step, and programming expertise and effort is needed to use and combine these tools.

An ideal system would directly generate an interface using example analysis queries.
Prior work~\cite{zhang2019mining} (\sysold) was a first step in this direction.
It modeled  interfaces as a visualization that renders and underlying query's result,
and widgets that syntactically transform the underlying query; the widgets defined the set of
all queries that the interface can express.
\sysold modeled each query as an abstract syntax tree (AST), aligned the trees, and 
extracted the subtrees that differ.  
It then grouped those differences and mapped each group to an interactive widget.  
In this way, it returned a set of widgets that expresses the input set of queries (and perhaps other queries).

Unfortunately, \sysold has several fundamental drawbacks.
Its output is limited to unordered set of widgets (e.g., \Cref{f:intro}(b))
because it doesn't consider how the query results are rendered. 
Thus, it cannot support interactions within visualizations (e.g., pan\&zoom, selection)
nor multiple visualizations in the same interface.
Further, it generates a flat mapping from syntactic differences to interactive widgets, however the flat mapping
cannot model hierarchical interface layouts nor nested widgets (such as tabs containing widgets).

This paper presents \sys, the first system to generate interactive multi-visualization
interfaces (such as \Cref{f:intro}(c)) from a small number of example queries.
To do so, we propose a new interface generation model that is based on schema matching.   
We first extend abstract syntax trees with four types of {\it choice nodes} to encode subtree variations between queries.  
Choice nodes directly correspond to grammar production rules.
For instance, the \texttt{ANY} choice node is used to choose one of its children---such as the two subqueries in \Cref{f:intro}(a))---and corresponds to an ordered choice production rule.

Each tree represents a subset of input queries, and we map the set of trees to an interface. 
Specifically, each tree's result table is mapped to a visualization, 
each choice node is mapped to a widget or visualization interaction,
and the tree structure is mapped to a hierarchical layout.  
By defining transform rules that merge, combine, and transform these trees,
we are able to search the space of tree structures that result in different candidate interface designs.  
Finally, we rank candidate interfaces by combining existing interface cost models~\cite{Gajos2010AutomaticallyGP,zhang2019mining,mackenzie1992extending} to estimate how easily the user can use the interface to express the sequence of input queries.  
\sys only requires access to a lightly annotated language grammar and database catalogue in order to determine valid mappings.

Informally, our technical problem is:
given an input sequence of queries, search the space of extended ASTs and interface mappings 
to identify the lowest cost interface.  
We use Monte Carlo Tree Search~\cite{Coulom2006EfficientSA,schadd2008single} (MCTS) to balance exploration of diverse tree structures with exploitation of good tree structures found so far, and
generate complex multi-view interfaces in seconds.  
We contribute:

\begin{myitemize} 
\item \sys, the first system to generate fully functional
  multi-view interfaces from example analysis queries.  
  The system is database agnostic, and only needs access to the query grammar,
  a database connection to execute queries, and the database catalogue. 

\item A novel model that unifies SQL query strings, 
  interactive visualizations, widgets, and interface layout. 
  The model enables us to formulate interface mapping in terms of schema mapping.

\item An evaluation that shows \sys expresses all data-related interactions in 
  Yi et al.'s ~\cite{Yi2007TowardAD}   visualization interaction taxonomy.  
  \sys shows that small difference in analysis queries can considerably change the interface design,
  and illustrates the importance of an automated interface generation tool.
  We further show 3 case studies that use real-world queries to improve the SDSS web search interface,
  reproduce Google's Covid-19 visualization, and show how to use queries to author a sales analysis dashboard.

\item A set of simple optimizations that reduce interface generation times from $30$s to a median of 6s.  
  We further find that \sys runtime scales linearly with the number of input queries.

\end{myitemize}

\stitle{Scope of this work:} \sys is designed to generate task-specific
analysis interfaces. It assumes a small sequence of coherent queries 
that represent a desired analysis,  and is not suitable for open-ended exploration queries that are often unrelated and seemingly random.  
Further, \sys is the first to show that end-to-end interface generation from queries {\it is even possible},
and we leave scalability and customizability of the generated interfaces to future work.

\section{Interface Generation Overview}\label{s:overview}

\sys transforms an input sequence of queries into an interactive interface in four steps: 
parsing queries into a generalization of abstract syntax trees (ASTs) that we call \difftrees, mapping the \difftrees to a candidate interface, estimating the interface's cost, and either returning the interface or transforming the \difftrees to generate a new candidate interface.  This section walks through these steps and introduces key concepts using a simple example.

\begin{figure}[b]
  \centering
  \includegraphics[width=\columnwidth]{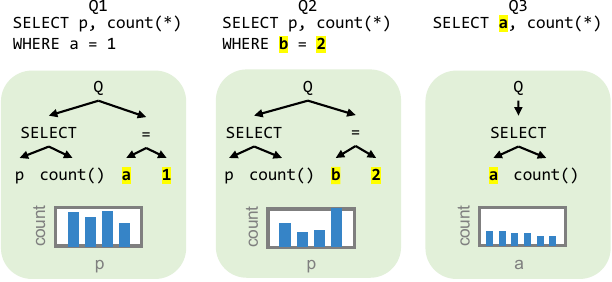}
  \caption{Example of three queries and their simplified ASTs.  A static interface would render one chart for each query.  }
  \label{f:log}
\end{figure}

\stitle{Static Interfaces: }
\Cref{f:log} lists three input queries on table \texttt{T} where attribute \texttt{p, a, b} are all integers.  \texttt{Q1} and \texttt{Q2} transform the predicate attribute and literal, and \texttt{Q3} selects \texttt{a} instead of \texttt{p}.  \sys first parses each query into their corresponding \difftree (simply a normal AST).  Each \difftree is rendered as a visualization; since the \difftrees are static, the interface for these three queries consists of three static charts.   For brevity, we omit the \texttt{FROM} and \texttt{GROUPBY} clauses and show simplified syntax trees.

\begin{figure}
  \centering
  \includegraphics[width=\columnwidth]{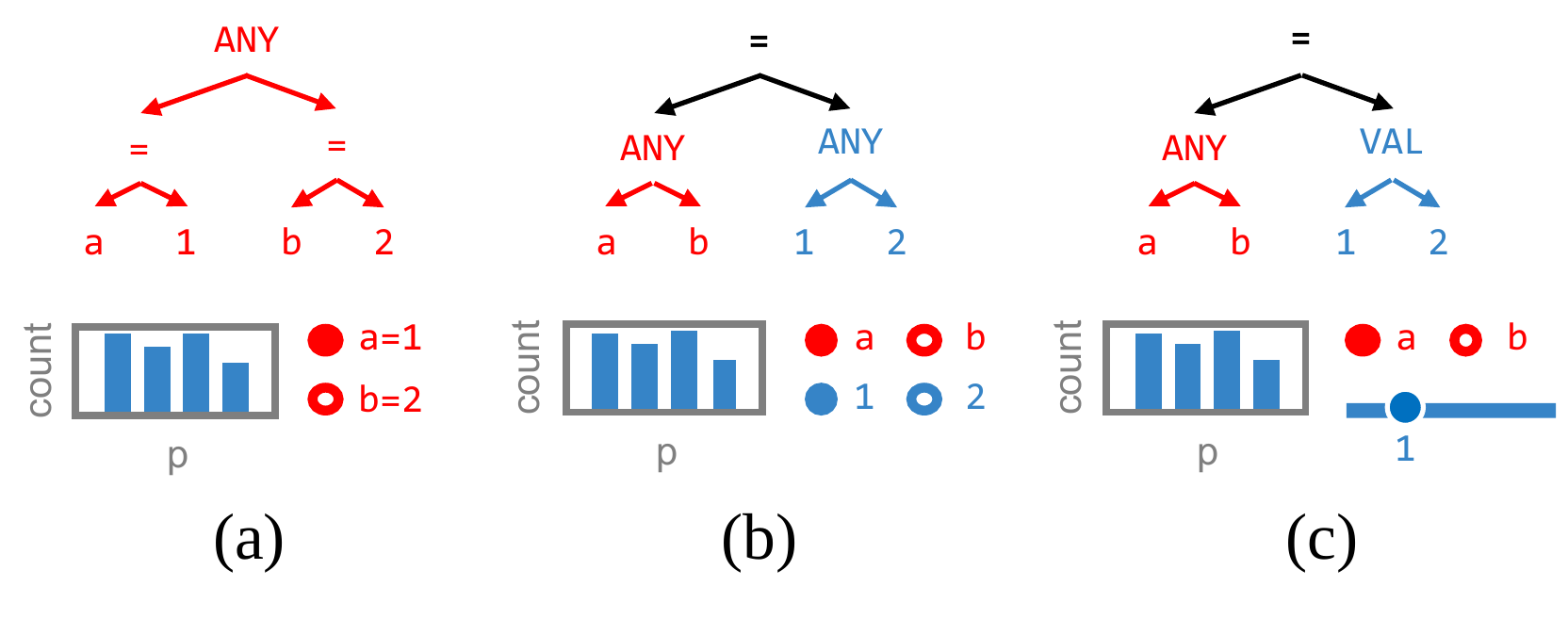}
  \caption{ Three examples of \difftrees for \texttt{Q1}, \texttt{Q2}, focusing on the subtree for the predicate.  The \texttt{ANY} choice node can choose one of its children.  (a) Each predicate can be chosen using a radio button that parameterizes the \texttt{ANY} node, (b) the left and right operands can be individually chosen using radio buttons, (c) the literal operand is generalized beyond the input values \texttt{1} and \texttt{2}.  }
  \label{f:q12}
\end{figure}

\stitle{Interactive Interfaces:}
Let us temporarily focus on the differing predicate in \texttt{Q1} and \texttt{Q2} to show how
different \difftrees structures can result in different interface designs.
For instance, \Cref{f:q12}(a) is rooted at an \texttt{ANY} node whose children are the two predicates.  \texttt{ANY} is a {\it Choice Node} that can choose one of its child subtrees.
In general, choice nodes encode subtree variations\footnote{Choice nodes can be viewed as generalizing parameterized expressions in SQL to parameterizing arbitrary syntax structures in a query. } that the user can control through the interface.
In the example, the \texttt{ANY} node is mapped to two radio buttons (other widgets such as a dropdown are valid as well), where clicking on the first
button would bind the \texttt{ANY} to its first child \texttt{a=1}.  
The \difftree output is visualized as a bar chart.

\ititle{Tree Transformations:} Note that both of \texttt{ANY}'s children are rooted at \texttt{=}, and can thus be pushed
above the \texttt{ANY} node.  This is an example of a {\it Tree Transformation Rule} that 
we describe in \Cref{s:search}.
The resulting \difftree in \Cref{f:q12}(b) shows two \texttt{ANY}
nodes that can independently choose the left and right operands.  This leads 
to an interface with two interactions (radio buttons), and also generalizes the interface beyond
the input queries.  For instance, the query can now express \texttt{SELECT p, count(*) WHERE b=1}.

\ititle{Schemas:} A single \difftree can be mapped to many interface designs, each with different widgets, visualization interactions, and layouts.
For instance, although \Cref{f:q12}(b) treated the second \texttt{ANY}'s children \texttt{1} and \texttt{2} as generic subtrees,
we can easily infer that they are numerically typed.  Further, the equality comparison
tells us their values are defined by the domains of attributes \texttt{a} and \texttt{b}.
Thus, with access to the database catalogue, we can infer that the second \texttt{ANY} node's
{\it Schema} is the union type of \texttt{a} and \texttt{b}, which are both numeric. 
We can then generalize the \texttt{ANY} to a \texttt{VAL} choice node that replaces itself
with the literal that it is bound to (\Cref{f:q12}(c)).
For instance, when the user changes the slider position to \texttt{5}, the value in bound to the \texttt{VAL} node, which resolves itself to \texttt{5}.
This lets us map the \texttt{ANY} node to a numeric slider that is initialized with the 
minimum and maximum of attribute \texttt{a} and \texttt{b}'s domains.
This was an example of generalization based on relaxing a choice node's schema (\Cref{s:types}), and 
a tree transformation rule that generalizes the \texttt{ANY} node's schema and replaces it with the \texttt{VAL} node.

\begin{figure}
  \centering
  \includegraphics[width=.8\columnwidth]{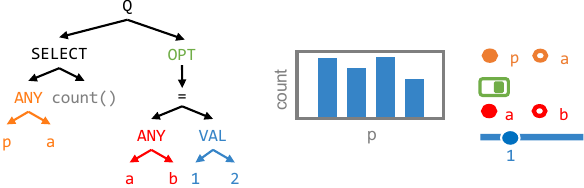}
  \caption{ A \difftree for \texttt{Q1-3} and a candidate interface.}
  \label{f:q123}
\end{figure}

\ititle{All Three Queries: } Now, let us add \texttt{Q3}.   
The simplest interface would be to partition the queries into two clusters, 
where  \texttt{Q3} is rendered as a static chart, 
and \texttt{Q1} and \texttt{Q2} is mapped to one of the interactive interfaces discussed so far.    
We can then choose to lay them out horizontally or vertically (\Cref{ss:layout}).
Another possibility is to merge all three queries into a single \difftree,
which would map to an interface with a single visualization.
\Cref{f:q123} illustrates one possible \difftree structure, 
where an \texttt{ANY} node in the \texttt{SELECT} clause chooses to project \texttt{p} or \texttt{a}.
This maps to an interface similar to \Cref{f:q12}(c), but with a radio  button to choose
the attribute to project and another toggle button to express optional status of the where clause.  
Naturally, which of these possible interface designs (or others not discussed here)
that should be generated and returned to the user depends on many factors, 
such as usability, layout, accessibility, and other factors that are difficult to quantify.  
Quantitative interface evaluation is an active area of research, and 
\Cref{ss:cost} presents the best practices used to develop this paper's cost function and its limitations.

\begin{figure}
  \centering
  \includegraphics[width=\columnwidth]{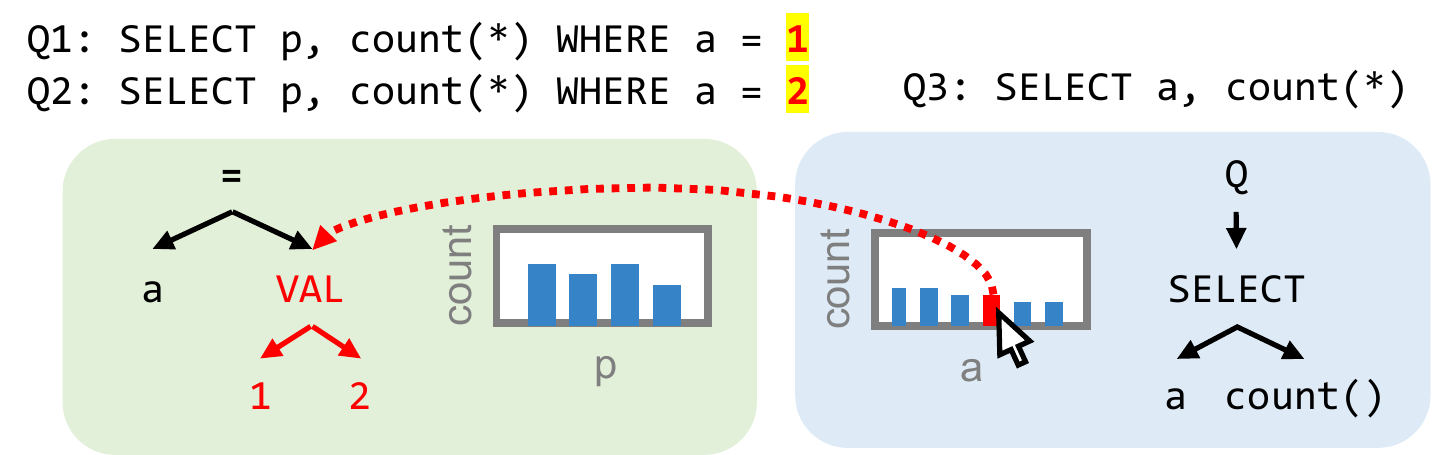}
  \caption{ Multi-view interface where clicking on the right-side chart updates the left chart.}
  \label{f:multiview}
\end{figure}

\stitle{Multi-view Interfaces}
\sys can also generate interactive multi-view interfaces.   
\Cref{f:multiview} illustrates a slightly different set of queries,
where the \texttt{Q1} and \texttt{Q2} only differ in the literal, and \texttt{Q3} remains the same.
Since the literal is compared to attribute \texttt{a}, an alternative to mapping 
the \texttt{VAL} node to a slider is to map it to a {\it visualization interaction}
in \texttt{Q3}'s bar chart.  Specifically, each bar is derived from \texttt{a} and \texttt{count(*)}
in \texttt{Q3}'s result.  Thus, clicking on a bar can also derive a valid
value in attribute \texttt{a}'s domain that can bind to the \texttt{VAL} node.

\begin{figure}
  \centering
  \includegraphics[width=\columnwidth]{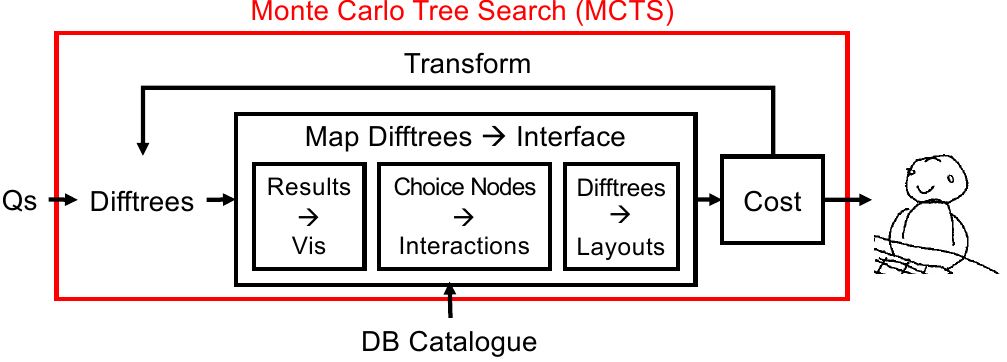}
  \caption{\sys interface generation pipeline.}
  \label{f:pipeline}
\end{figure}

\stitle{Summary and Generation Pipeline:}
To summarize, \sys generates interfaces in a four-step process.
We first parse the input query sequence $Q$ into \difftrees, and map the \difftrees
into an interface.  An interface mapping $\mathbb{I}=(\mathbb{V},\mathbb{M},\mathbb{L})$ is defined by mapping each \difftree
result to a visualization ($\mathbb{V}$), choice nodes to interactions (widget or visualization interaction) ($M$),
and a layout tree ($\mathbb{L}$).  A cost function $C(\mathbb{I}, \mathbb{Q})$ evaluates the interface and either returns the interface or 
choose a valid transformation to apply to the \difftrees.  
Informally, given queries $Q$ and cost function $C$, our problem is to return the lowest cost interface $\mathbb{I}$ that can express all queries in the log.
  The formal statement is presented in \Cref{s:search}.

We solve this problem using Monte Carlo Tree Search~\cite{Browne2012ASO} (MCTS), a search algorithm for learning good game-playing strategies and famously used in AlphaGo~\cite{Silver2016MasteringTG}.
It balances exploitation of good explored states (\difftree structures), and exploration of new states.
We describe our search procedure and optimizations in \Cref{ss:mcts}.

\section{Difftrees}
\label{s:difftree}

\difftrees extend ASTs with {\it Choice Nodes} that encode the structural differences between the queries,
and are the bridge from input queries to the output interface.
Specifically, an interface $\mathbb{I}=(\mathbb{V},\mathbb{M},\mathbb{L})$ is defined by mapping each \difftree result to a visualization ($\mathbb{V}$), choice nodes to interactions ($\mathbb{M}$), and a layout tree ($\mathbb{L}$). 
Candidate visualization and interaction mappings are based on schema matching between
the \difftree result and visualization schemas, and choice node and interaction schemas respectively.
This section defines \difftrees and how their result and choice node schemas are inferred.  
The next section will present visualization and interaction schemas and the formal mapping procedure.

\subsection{Difftree and Choice Nodes}
\label{ss:choicenodes}

A \difftree $\Delta$ compactly represents a set of expressible ASTs
$\{\Delta\} = \{\delta_1, \delta_2, ...\}$, where $\delta_i$ is an AST.
It extends ASTs with {\it Choice Nodes}--\texttt{ANY}, \texttt{VAL}, \texttt{MULTI}, and \texttt{SUBSET}--that correspond to production rules in a PEG grammar.
This supports arbitrarily complex subtrees, yet can still be analyzed because the set of variations is predefined and finite.  
Finally, \difftrees guarantee that any expressible AST is syntactically correct.
We describe \difftree node types below, the set of ASTs they express, and how each choice node resolves to an AST subtree when bound to a set of parameters.  

\begin{myitemize}
  
  \item \stitle{\texttt{ANY(c1,..,ck)}} can choose one of its $k$ children, akin to the production rule \texttt{ANY$\to$c1|..|ck}.  
    When bound to an index $i\in[1,k]$, it resolves to $c_i$. 
    For instance, binding $1$ to the \texttt{ANY} in \Cref{f:q12}(a) will resolve it to \texttt{a=1}.
    A special case is when \texttt{ANY} has two children, where one is an empty subtree. 
    We call this \texttt{OPT} for optional, and is useful for mapping to binary interactions such as toggles.
    This node expresses the ASTs: $\cup_{i\in[1,k]}\{c_i\}$.

  \item \stitle{\texttt{VAL(c1,..,ck)}} represents a literal that matches a regex pattern in a grammar. 
   Its children are all literals
    and its value domain is defined by the union of its children's types. In practice, it is a pass-through node
    that resolves to any value it is bound to. 
    For instance, the \texttt{VAL} in \Cref{f:q12}(c) resolves to the slider's value. 
    The next subsection describes types in more detail.  
    Let $c_i.d$ refer to a child node's domain, then this node expresses $\cup_{i\in[1,k]} c_i.d$.

  \item \stitle{\texttt{MULTI[sep](c)}} represents lists that express e.g., project lists, group-by lists, and conjunctions.
    It expresses the production rule \texttt{MULTI$\to$c (sep c)*}.
    It repeats its child \texttt{c} one or more times, where its child may also be a \difftree.  
    When it is bound to a list of parameterizations $[p_1,..,p_k]$, it passes each parameterization $p_i$ to its child 
    and concatenates their resolved ASTs using \texttt{sep}.
    This node expresses the ASTs: $\{sep\left(t\right)  | t\in \{c\}^k \land k\in\mathbb{N}_0\}$ that correspond to an arbitrary number of cross products between all elements of $\{c\}$.
    
  \item \stitle{\texttt{SUBSET[sep](c1,..,ck)}} represents the production rule \texttt{SUBSET$\to$\\c1?..ck?} with \texttt{sep} as the separator. 
   Bind a set of indices between $1$ and $k$ to it will resolve 
    to the corresponding subset of its children, concatenated with the separator \texttt{sep}.
    The node expresses the ASTs: $\{ sep\left(t\right) | t\subseteq \{c1, c2,... ck\} \}$ that corresponds to all possible subsets of the children.

  \item \stitle{Non-choice Nodes: } Finally, let \texttt{N(c1,..,ck)} be a non-choice node.  
If $N$ is a leaf node, it expresses the ASTs: $\{N\}$.
Otherwise, it expresses $\{N(t) | t\in\{c_1\}\times\ldots\times\{c_k\}\}$.
Let $c_{s_1},\ldots,c_{s_n}$ be the subset of children whose subtrees contain one or more choice nodes.
When $N$ is bound to a list of parameterizations $[p_1,\ldots,p_n]$, it passes each $p_i$ to
child $c_{s_i}$.

\end{myitemize}

\subsection{Schemas}
\label{s:types}

Although \sys does not reproduce the database front-end's type checking,
it infers type and schema information to map \difftrees to an interface.
Choice nodes and their ancestors are annotated with schemas, and all other nodes are annotated with types.  
We distinguish these because only choice nodes (or their ancestors) are mapped to interactions.
We use {\it Result Schema} to refer to the schema of the \difftree's result,
and {\it Node Schema} to refer to the schema of a \difftree node.

\subsubsection{Types}
\label{ss:type}

All non-choice nodes that are not ancestors of a choice node are annotated with type information.  
A type defines a  domain of values~\cite{aho1986compilers} that a node can express.   For simplicity we describe a trivial type hierarchy of primitive types:  AST$\to$str$\to$num.  num specializes str, and str specializes AST; AST expresses any abstract syntax tree.  Further, each database attribute \texttt{a} itself represents an {\it Attribute Type} and specializes a primitive type's domain to \texttt{a}'s domain.  In general, internal nodes are of type \texttt{AST}, while leaf nodes have more specialized types.  We say that a type $t_1$ is {\it compatible} with $t_2$ if its domain is a subset of $t_2$'s domain.

\stitle{Initialization: }
We initialize the leaf node's types by using lightweight grammar annotations and the database catalogue.  
Specifically, we annotate production rules that resolve to str, num with its corresponding type.
Also, we infer the type of a function call based on it return type in the catalogue.

\stitle{Inference: }
When possible, it is useful to specialize a primitive type to an attribute type.  For instance, given $a = 1$, we would like to infer that $1$ has type $a$.
To do so, we first lookup attribute names in the catalogue to determine its fully qualified attribute name and domain.  
We use a simple heuristic based on equality comparison expressions of the form \texttt{attr = val}, and assign \texttt{val}'s type as \texttt{attr}.
Finally, we define the union of two types $T_1 \cup T_2$ as their least common ancestor in the type hierarchy. For instance, \texttt{str=num}$\cup$\texttt{str}, while \texttt{num=num}$\cup$\texttt{num}.

\begin{figure}[bt]
  \centering
  \includegraphics[width=\columnwidth]{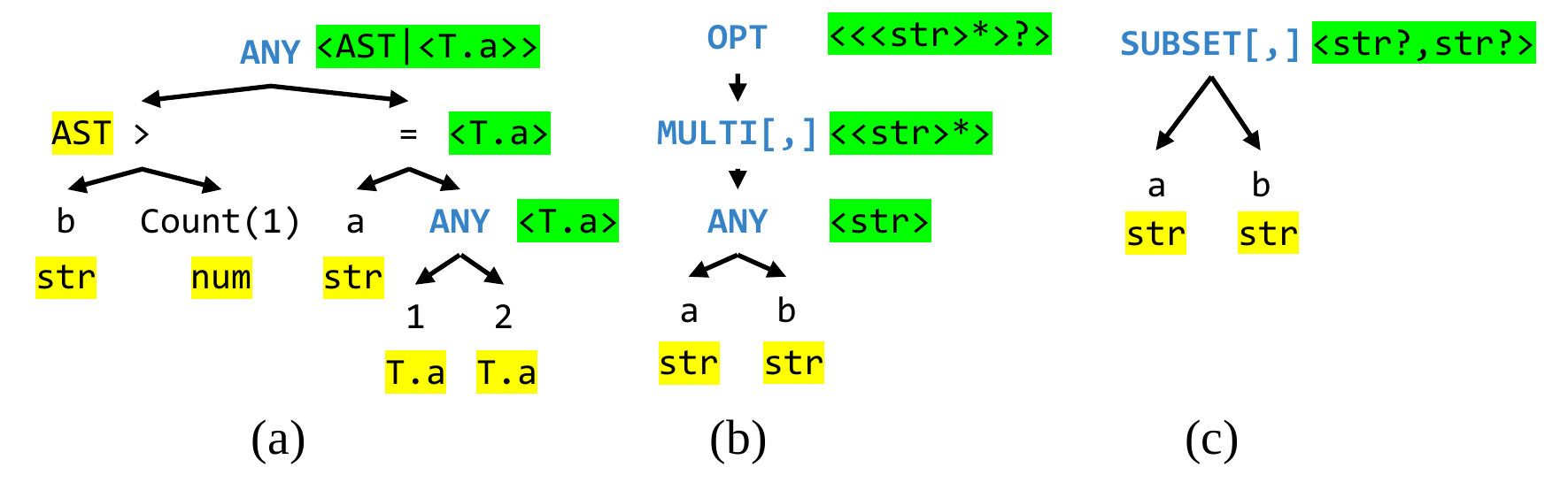}
  \caption{Example \difftrees of the four types of choice nodes, with \highlight{type} and \ghighlight{schema} annotations. } 
  \label{f:inference}
\end{figure}

\begin{example}[Node Types]
  \Cref{f:inference}(a) illustrates examples of type inference in \highlight{yellow}. 
  \texttt{b}, and \texttt{a} are \texttt{str} types that refer to attribute names (note, they are not attribute types themselves). The catalogue lists \texttt{count()} as type \texttt{num}.  $1$, $2$ are \texttt{num} types, however they are compared with \texttt{a} so their types are specialized to the fully qualified \texttt{T.a}.
\end{example}

\subsubsection{Result Schemas} 
\label{ss:resultschema}

The result schema is defined for a \difftree $\Delta$ based on the set of ASTs $\{\delta_1, \delta_2, ...\}$ that it expresses.  Specifically, let $s(\delta)=<a^\delta_1:t^\delta_1,..,a^\delta_n:t^\delta_n>$ be the standard result schema of an AST $\delta$ without choice nodes, where $a_i$ is the attribute name, and $t_i$ is its type.  $\Delta$'s result schema is well defined if all $s(\delta\in\{\delta_1, \delta_2, ...\})$ are union compatible.  In this case, its result schema is $<a_1:t_1,\ldots,a_n:t_n>$ where $a_i=\{a^\delta_i | \delta\in\{\delta_1, \delta_2, ...\}\}$, and $t_i=\cup_{\delta\in\{\delta_1, \delta_2, ...\}} t^\delta_i$.  

In short, each attribute name is a concatenation of the unique attribute names, and each type is the least compatible type across all expressible ASTs.  For example, the result schema in \Cref{f:q123} is $<\{T.a\cup T.p\}, num>$.  If the schemas are not union compatible, the result schema is undefined.

\subsubsection{Node Schemas}
\label{ss:nodeschema}

A choice node or an ancestor is a {\it Dynamic Node}, and all other nodes are {\it Static}.
Dynamic nodes are annotated with node schemas that describe the structural variation that they express.   
Schema $<e_1,\ldots,e_n>$ is a list of type expressions, where each expression $e_i$ is a set operators $\texttt{\{|, ?, *\}}$ over types and schemas.  
\texttt{\{|, ?, *\}} have regular expression semantics where $|$ is the \textit{or} relation 
(\texttt{ANY}), $?$ is the \textit{existential} relation (\texttt{OPT}, \texttt{SUBSET}), and $*$ is \textit{repetition} (\texttt{MULTI}).

\stitle{Node Schema Inference:}
Let $N(c_1,\ldots,c_n)$ denote a dynamic node and its $n$ children, where its children may be dynamic or static nodes.  
Further, let $T(N)$ refer to the node $N$'s type if it is static, and its schema if it is dynamic.  
We now define schema inference rules that define $T(N)$ for dynamic nodes.
Note that schemas may be nested in order to accomodate nested interfaces, such as tabs.

\begin{myitemize}
\item \stitle{\texttt{ANY(c1,..,cn)}} considers two conditions.  
If all children are static, then $T(ANY) = <\cup_{i\in[1,n]}T(c_i)>$.  Its schema is the least compatible type of its child types.
Otherwise, $<T(c_1) | \ldots | T(c_n)>$ is the \texttt{OR} of its child schemas.

\item \stitle{\texttt{OPT(c)}:} $<T(c)?>$.  

\item \stitle{\texttt{MULTI[sep](c)}:} $<T(c)*>$, where $*$ denotes $0+$ repetitions.

\item \stitle{\texttt{SUBSET[sep](c1,..,cn)}:} $<T(c_1)?,\ldots,T(c_n)?>$. 

\item \stitle{\texttt{AST(c1,..,cn)}} considers two conditions.
A static node (\Cref{ss:type}) has type \texttt{AST}.
Otherwise, its schema is the cross product of its dynamic children's schemas $<T(c_{s_1}),\ldots,T(c_{s_k})>$,
where $c_{s_1},\ldots,c_{s_k}$ are its dynamic children.

\end{myitemize}

\begin{example}
  \Cref{f:inference} annotates \difftrees with node schemas in \ghighlight{green}.    In \Cref{f:inference}(a), the bottom \texttt{ANY} only has static children, so its schema is the union of its child types.  The schema of the AST node \texttt{=} is the cross product of its dynamic children's schemas, which is simply $<T.A>$.  Finally, the top \texttt{ANY} expresses its left or right child, so has a nested schema $<AST|<T.a>>$.  
  \Cref{f:inference}(b) illustrates nested schemas, where \texttt{MULTI} applies \texttt{*} to its child schema, and similarly \texttt{OPT} applies \texttt{?}.  (c) is an example of \texttt{SUBSET}.

\end{example}

\subsubsection{Query Bindings}

We wish to guarantee that the generated interface can express all of the input queries (and possibly more).  
Schema information is unfortunately insufficient to ensure this guarantee.    For instance, suppose a \texttt{VAL}
node has type $T.a$ with two child literals \textit{1} and \textit{100} from queries $q_1$ and $q_2$.
In addition, a bar chart's x-axis renders $T.a$.   Naively, one might expect that clicking on bars in the bar chart can express $T.a$ values, and thus can be bound to the \texttt{VAL} node.  However, it is possible that the query generating the bar chart filters out all records where $T.a=100$, and thus cannot express $q_2$.

Thus, we also derive the set of {\it Query Bindings} needed for each dynamic node in order to express all of the input queries.
A query binding is a tuple of values consistent with the node's schema.
This is done by tracking the binding needed for the \difftrees to express each input query and unioning the bindings on a per-node basis.

\begin{example}
  Consider \Cref{f:inference}(b) and two input queries \texttt{a,a} and \texttt{b}.  
  The \difftree bindings are \texttt{\{Multi:[\{ANY:1\}, \{ANY:1\}]\}} and \texttt{\{Multi:[\{ANY:2\}]\}}.
  The bindings for each choice node are the union across query bindings.  
  For instance, the bindings for \texttt{MULTI} are \texttt{\{[\{ANY:1\}, \{ANY:1\}],[\{ANY:2\}]  \}},
  and those for \texttt{ANY} are \texttt{\{1,2\}}.
\end{example}

We will refer to Query Bindings when determining whether a candidate interaction mapping is {\it safe}, described next.

\section{Interface Mapping}
\label{s:mapping}

\sys generates candidate visualization ($\mathbb{V}$), interaction ($\mathbb{M}$), and layout ($\mathbb{L}$) mappings in order to generate an interface $\mathbb{I}=(\mathbb{V},\mathbb{M},\mathbb{L})$.  These candidates define the search space that \sys explores in \Cref{s:search}.  
Visualization and interaction mappings are grounded in schema matching and seeks to ensure safety (that the interface can express all input queries).   \sys is extensible, in that developers can add new visualization types, interaction templates, as well as different types of layouts beyond those used in our prototype.
Finally, we will formally present the interface generation problem.

\subsection{Visualization Mapping $\mathbb{V}$}
\label{ss:vismap}

$\mathbb{V}$ defines the set of mappings from each \difftree to the visualization that renders its results.  
Although there are numerous visualization recommendation algorithms, such as ShowMe~\cite{Mackinlay2007ShowMA}, Draco~\cite{Moritz2019FormalizingVD}, and Deepeye~\cite{luo2018deepeye}, each is focused on an individual output chart.
In contrast, \sys generates multi-visualization interfaces and needs to take the entire interface into consideration. 
Specifically, some visualization types, although not individually optimal, may enable visualization interactions
that improve the overall interface.   For this reason, we use a simple set of heuristics to map a \difftree to a visualization.

\stitle{Visualizations as Schemas:}
A visualization renders records from an input table as marks (points, bars) in a chart.
Each visualization encodes data attributes using a set of visual variables~\cite{bertin1983semiology}, such as x, y, size, and color, and makes different assumptions about the data types mapped to those visual variables.

As such, we model each visualization type using a {\it Visualization Schema} $<a_i:t_i,\ldots>$, where $a_i$ is the name of a visual variable, and $t_i$ is either Quantitative (Q) or Categorical (C) type.
For instance, a bar chart renders categorical values along the x axis, quantitative values along the y axis, and optionally renders categorical values as the bar color.  

In addition, a visualization may enforce functional dependency (FD) constraints over the input data.  For instance, bar charts assume that x and color functionally determine the y value.  This can be inferred if data is a group-by query since the grouping attributes determine the aggregate values, or if the attributes mapped to x and color are unique. \Cref{tab:vis} summarizes the schemas and constraints.

\begin{table}\small
  \begin{tabular}{@{}rll@{}}
  \textbf{Vis} & \textbf{Schema and FDs} & Interactions\\  \midrule
  \textbf{Table} & any schema & Click\\ \midrule
  \textbf{Point} & $<$\texttt{x:Q|C, y:Q,shape:C?, } & Click, Multi-click, \\
          &  \texttt{ size:C?, color:C?}$>$ & Brush-x/y/xy, Pan, Zoom \\ \midrule
  \textbf{Bar}   & $<$\texttt{x:C, y:Q, color:C?>}$$ & Click, Multi-click\\
                 & (x, color)$\to$y & Brush-x \\\midrule
\textbf{Line}  & $<$\texttt{x:Q|C, y:Q, shape:C?,} & Click, Pan, Zoom \\ 
               & \texttt{ size:C?, color:C?}$>$ &  \\
                & (x, shape, size, color)$\to$y & \\ \bottomrule
\end{tabular}
\caption{Visualization schemas, FD constraints, and supported interactions. \texttt{Q} and \texttt{C} stand for quantitative (numeric) and categorical types.}
\label{tab:vis}
\end{table}

\stitle{Visualization Mappings:}
A visualization $V$ can render the result of a \difftree $\Delta$ if there is a valid mapping from $\Delta$'s result schema $S_\Delta$ and the visualization schema $S_V$ such that (1) every data attribute $d$ is mapped to a visual attribute $v$, (2) each visual attribute is mapped to at most once, (3) every non-optional visual variable is mapped to, and (4) $d$'s type in the result schema is compatible with $v$'s type in the visualization schema.
We define compatibility as follows: \texttt{str}, and \texttt{num} attributes whose cardinality is below 20, are compatible with categorical visual attributes, and \texttt{num} attributes are compatible with quantitative visual attributes.
Finally, we check that the \difftree result satisfies the visualization's constraints based on the query structure (whether it is a group-by) and database schema constraints.

\begin{example}
  Consider \texttt{Q1} in \Cref{f:log}, which groups by \texttt{p} and computes \texttt{count(*)}.  We can infer that \texttt{p} determines \texttt{count}, use the database statistics to estimate that \texttt{p}'s cardinality is below 20, and thus infer that it can be mapped to a quantitative or categorical visual attribute.  The bar chart mapping $\{ p\to x, count\to y \}$ satisfies the type compatibility and functional dependency constraints, and maps all attributes in the result schema, and has mappings to the required visual attributes.  
\end{example}

\stitle{Candidate Generation:} 
We generate all valid mappings for a \difftree by iterating through each visualization type, and generating all permutations of the result schema that result in a valid mapping to the visualization schema.

\subsection{Interaction Mapping $\mathbb{M}$}

$\mathbb{M}$ defines the set of interaction mappings from dynamic nodes to widgets or visualization interactions (collectively called interactions).
An interaction mapping $\delta\to I$ from a dynamic node $\delta$ to interaction $I$ means that
when the user manipulates the interaction, it generates a stream of event tuples whose values bind to $\delta$ (\Cref{ss:choicenodes} describes node bindings).
The goal is to generate $\mathbb{M}$ such that there is a binding for every choice node in the \difftrees.  

At a high level, \sys manages a library of interaction templates, and checks which templates are valid for a given dynamic node.
If valid, \sys instantiates the interaction with the dynamic node's information, and binds its manipulation event stream to those choice nodes.  
This subsection first models interactions as schemas and domain constraints, 
and then defines valid and safe interaction mappings.

\subsubsection{Interaction Model}
\label{ss:interaction}

\sys manages an extensible library of interaction templates (widgets and visualization interactions).
An interaction template defines a schema that is used to identify candidate choice node mappings, and optional constraints that are applied to the choice nodes' query bindings.  

An interaction mapping $\delta\to I$ is {\it valid} if there is a schema match from the dynamic node's schema $S_\delta$ to the interaction's schema $S_I$, and $\delta$'s query bindings satisfy the interaction's constraints (if any).  Specifically, a schema match exists if (1) $S_\delta$ and $S_I$ have the same number of type expressions, and (2) each type expression $e^\delta_i$ in $S_\delta$ is compatible with the corresponding expression $e^I_i$ in $S_I$.
\Cref{tab:widgets} lists example widget schemas and constraints.

\begin{figure}
  \centering
  \includegraphics[width=\columnwidth]{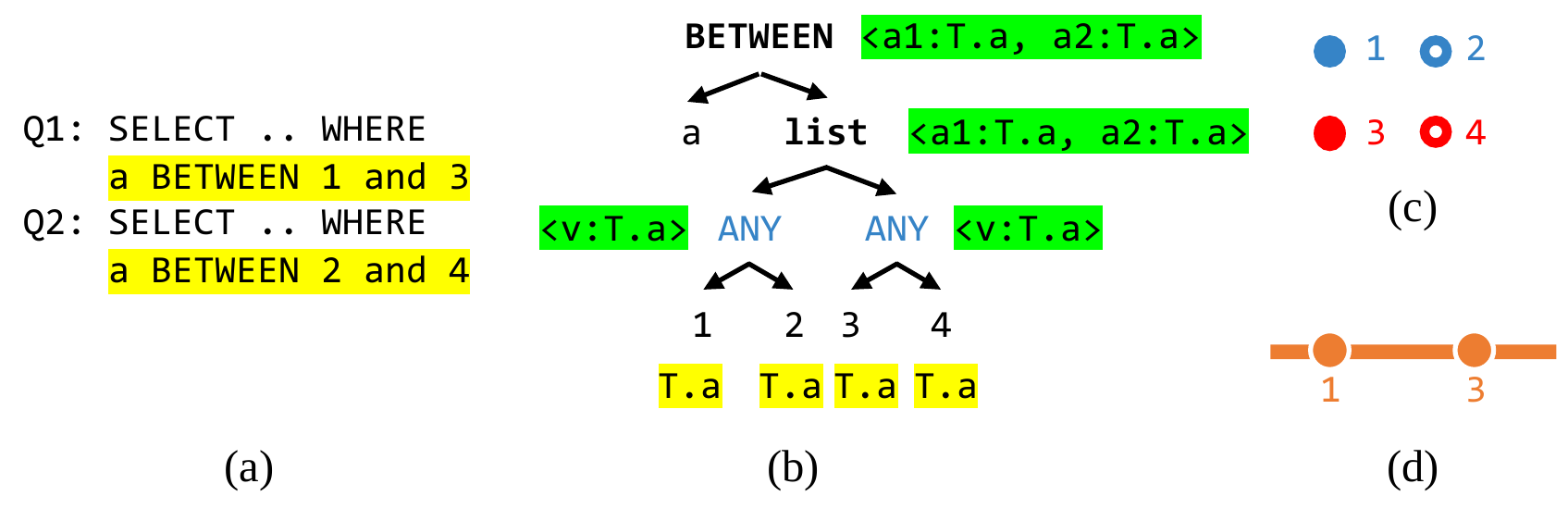}
  \caption{Annotated \difftree for highlighted portion of \texttt{Q1} and \texttt{Q2}.  (b, c) are candidate interaction mappings for the \texttt{BETWEEN} or \texttt{list} dynamic nodes.}
  \label{f:twochoice}
 \end{figure}

\begin{example}
  Consider the \difftree in \Cref{f:twochoice}.  
  A radio list has schema \texttt{<\_>}, where \texttt{\_} matches any schema or type expression, and is compatible with each \texttt{ANY} node's schema.  
  The schema mapping for each \texttt{ANY} node would be $(ANY.v\to radio.v)$.

  Alternatively, A range slider has schema \texttt{<s:num, e:num>}. 
  The \texttt{list} node in \Cref{f:twochoice}(b) has schema \texttt{<a1:T.a,a2:T.a>}, which is compatible with the range slider's schema because $T.a$ is numeric.  
  The schema mapping is $(a1\to s, a2\to e)$.  
  Note that since \texttt{list} is not a choice node, the event tuples generated by the range slider
  that are bound to the node will be routed to its child \texttt{ANY} nodes.  Thus, the two \texttt{ANY}
  nodes are bound by this interaction mapping.
\end{example}

The second criteria for a valid mapping is that $\delta$'s query bindings satisfy the interaction's constraints.
For instance, the range slider has the constraint that the start value $s$ is $\le$ to the end value $e$.
We can see that the query bindings for the \texttt{list} node in \Cref{f:twochoice} are $(1,3)$, $(2,4)$, and 
satisfy the constraints.  Thus, it is valid to map the \texttt{list} node to the range slider.

Finally, a mapping is {\it safe} if there exist user manipulations to produce event tuples for each of the dynamic nodes' query bindings.   For widgets, safety is ensured by construction because each widget is initialized with the dynamic node's  query bindings.  Safety is not always guaranteed for visualization interactions, and we discuss this below.

\stitle{Widgets:}
\sys is prepopulated with a library of common widgets, including button, radio list, checkbox list, dropdown, slider, range slider, adder, and textbox.  For space reasons, \Cref{tab:widgets} lists a subset of their schemas and constraints.

\begin{table}\centering\small
\begin{tabular}{rll}
  \textbf{Widgets} & \textbf{Schema} & \textbf{Constraint} \\\hline
  Radio, Dropdown, Textbox  & \texttt{<v:\_>} &  \\
  Toggle & \texttt{<v:\_?>} & \\
  Checkbox & \texttt{<v:\_*>} & \\
  Slider & \texttt{<v:num>} & \\
  RangeSlider & \texttt{<s:num,e:num>} & \texttt{s$\le$e}\\
\end{tabular}
\caption{Example widget schemas and constraints.  \texttt{\_} matches any schema or type expression.}
\label{tab:widgets}
\end{table}

\begin{figure}
  \centering
  \includegraphics[width=\columnwidth]{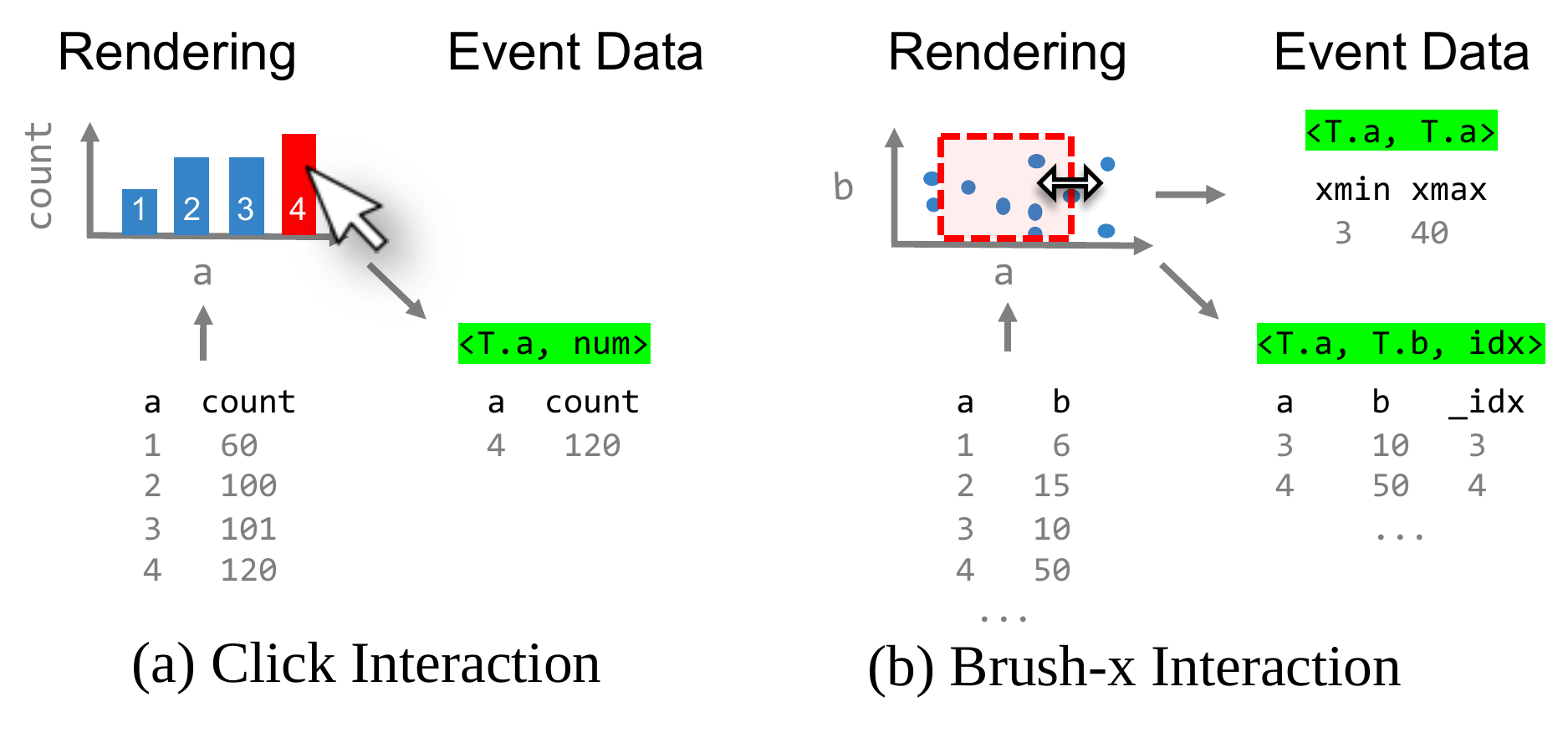}
  \caption{Examples of visualization interactions and their event data.  A visualization maps each input record to a mark in the chart. User manipulations generate one or more event streams.  Event schemas are \ghighlight{in green}. }
\label{f:interaction}
\end{figure}

\stitle{Visualization Interactions:}
A visualization is modeled as a one-to-one projection of input records to marks rendered on the screen.
There are three concerns when modeling interactions in visualizations:
(1) each visualization type can support multiple interaction types (e.g., click, brush, pan),
(2) each interaction can generate multiple event streams during user manipulations, and 
(3) the schemas of the event data depend on the visualization's own mapping (\Cref{ss:vismap}).
We illustrate these concerns using the bar chart and scatterplot in \Cref{f:interaction}:

\begin{example}
  The bar chart renders its input four records using the visualization mapping $(a\to x, count\to y)$.
  One of the interactions that a bar chart supports is click interactions.  
  For instance, if the user clicks the fourth bar, it corresponds to selecting the fourth input record,
  and thus the event stream emits $(4, 120)$ with the schema \texttt{<T.a, num>}.

  The scatterplot renders its input records using the mapping $(a\to x,b\to y)$.
  It illustrates a 1-D brush interaction (along the x-axis) that emits two event streams.
  The first stream represents the minimum and maximum bounds of the selection box and has schema \texttt{<T.a,T.a>},
  while the second stream represents the set of selected records, and thus has the same schema as
  the input data.  \sys internally tracks the index of each record, which is useful for binding \texttt{ANY} nodes.

  Interaction schemas can depend on the visualization mapping. 
  If the scatterplot used the mapping $(b\to y,a\to x)$,  
  then the first stream's schema would instead be \texttt{<T.b, T.b>}.
\end{example}

To this end, each visualization type defines a set of interactions and the event stream schemas for each interaction.  The schemas are specified in terms of the visualization's visual attributes, and \sys uses the visualization mapping to automatically translate them to be in terms of \difftree's result schema.  Thus, each visualization mapping corresponds to a set of interaction event schemas that are candidates for interaction mapping.    \Cref{tab:vis} summarizes the interactions each visualization type supports.

\subsubsection{Visualization Interaction Safety}

Visualization interactions introduce a unique safety concern because their input data is based on the result of a \difftree.  In contrast to widgets, whose domains are initialized based on query bindings in the \difftree and thus guaranteed to be safe, the values that a visualization interaction can express depends on the contents of its input data.    Concretely, consider the bar chart's click interaction in \Cref{f:interaction}.  The output of its \difftree $\Delta$ contains four records, thus the click interaction can express the $T.a$ values: $1,2,3,4$.   Suppose there is a choice node \texttt{VAL(4,5)} with schema \texttt{<num>}.   Based on the above rules, it is valid to map this choice node to the click interaction because their schemas match.  However, this specific chart {\it cannot express the query binding $5$}!

We use a simple heuristic to check safety. Given a candidate interaction mapping to an interaction in visualization $V$, we can check $V$'s visualization mapping $\Delta\to V$.  Since we know the subset of input queries that $\Delta$ expresses, we can (logically) instantiate the visualization with each query's result table, and check the subset of query bindings that the interaction can express.   If there exists an input query that can express every query binding, then the interaction mapping is safe.  This heuristic appears effective in practice, however as we see in the runtime experiments, checking safety for every candidate mapping degrades \sys runtime when there are many input queries.   We leave optimizations to future work.

\subsection{Layout Mapping $\mathbb{L}$}
\label{ss:layout}

After visualization and interaction mapping, we can mark the dynamic nodes in the \difftrees that correspond to widgets on the screen.  $\mathbb{L}$ defines a layout tree over the visualizations and widgets, where each layout node either horizontally (\texttt{H}) or vertically (\texttt{V}) lays out its child elements.  

Let a \difftree node that has been mapped to a widget be called a {\it Widget Node}.  For a given \difftree $\Delta$, we first create a layout tree $W_\Delta$ for its widgets: each widget is a leaf node, and we create a layout node for the least common ancestor of every pair of widget nodes in the \difftree.  $\Delta$'s layout tree $L_\Delta$ is a layout node whose children are $W_\Delta$ and the \difftree's visualization.   The final layout tree $\mathbb{L}$ is a root layout node whose children is each \difftree's layout tree. 
Note that layout is currently best effort, users are free to change the widget positioning themselves.

One point of note is that some widgets may themselves be layout nodes.  For instance, a radio list, toggle, or tab list may be used to choose different sub-interfaces, but also require space to render the widget themselves.  Typically, these ``layout widgets'' will be mapped to choice nodes that have descendant choice nodes, and thus takes the place of a layout node.  

Finally, \sys also uses the layout tree to estimate the bounding box of every node in the tree.    This is used during interface cost estimation to assess the amount of effort that the user must move between widgets and visualizations in order to express a given query, and penalize interfaces that exceed an optional screen size.  To do so, we also estimate text and widget sizes based on their initialization parameters.

\begin{figure}
  \centering
  \includegraphics[width=.85\columnwidth]{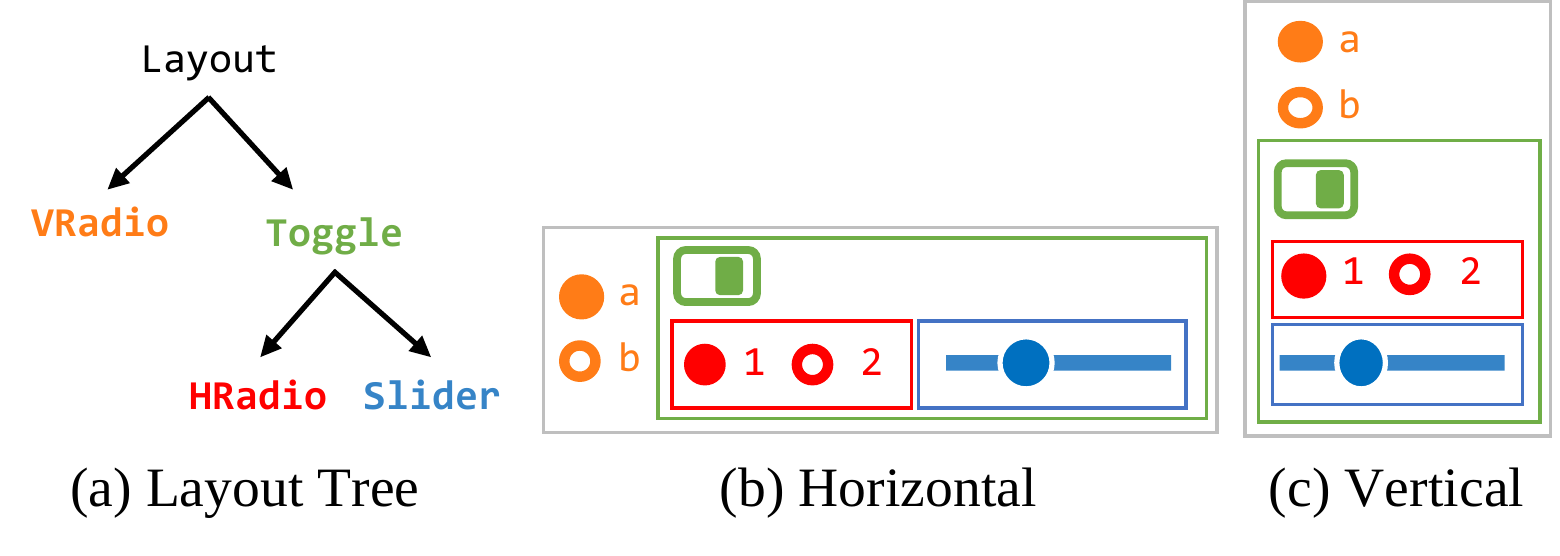}
  \caption{Layout tree and their bounding boxes when the layout nodes (\texttt{Layout}, \texttt{Toggle}) are horizontal or vertical.}
  \label{f:widgettree}
 \end{figure}

\begin{example}
  \Cref{f:widgettree} illustrates a layout tree with three leaf widgets, the toggle layout widget, and a root layout node.  Vradio (Hradio) renders its elements vertically (horizontally), and the toggle button is always in the top left above its children.  \Cref{f:widgettree}(a) and (b) illustrate the layout and bounding boxes if the layout nodes were horizontal or vertical, respectively.
\end{example}

\section{Cost Model}
\label{ss:cost}

The space of possible interfaces is very large, and a cost model is needed to estimate the quality of a candidate interface mapping.  This is deep and long standing problem in HCI research, and a common measure is based on the expected time the user will take to perform the manipulations needed to complete a set of tasks using an interface~\cite{Sears1993LayoutAA, Gajos2010AutomaticallyGP,Card1983ThePO,Kieras1994GOMSMO}. 
\sys uses a simple cost model $C(\mathbb{I},Q)=C_U(\mathbb{I},Q) + C_L(\mathbb{I})$ that combines usability and layout characteristics of an interface.
For usability, we estimate $C_U$ as the time needed for the user to express the sequence of input queries using the interface.  
For layout, $C_L$ accounts for the interface size.

\stitle{Usability:}
We measure usability based on SUPPLE~\cite{Gajos2010AutomaticallyGP}, 
which models the interface cost $C_U(\mathbb{I},Q) = C_m(\mathbb{I}, Q) + C_{nav}(\mathbb{I},Q)$ as the time to manipulate each widget (or visualization) $C_{m}$ and the time to navigate between interactions $c_{nav}$.

Manipulation cost $C_m(w)$ for a widget $w$ is modeled as a second order polynomial 
$C_m(w) = a_0 + a_1|w.d| + a_2|w.d|^2$, where $|w.d|$ is the size of the widget's domain.  Widgets that enumerate options (e.g., radio, dropdown, and checkboxes) define $|w.d|$ as the number of options; other widgets set $|w.d|=0$.  The manipulation cost for the interface $C_m(\mathbb{I})$ is the total cost of manipulating the widgets needed to express each input query.
Our prototype uses parameters fit to widget interaction traces in prior work~\cite{Chen2020MonteCT,zhang2019mining}, and sets visualization interaction costs to low constants to encourage choosing them.

The navigation cost $C_{nav}$, proposed in SUPPLE~\cite{Gajos2010AutomaticallyGP}, is based on Fitts' law~\cite{Sears1993LayoutAA,mackenzie1992extending}, a model of human movement widely used in HCI and ergonomics.  It states that the time to move to a target area increases with its distance $D$ and inversely to its width $W$ along the axis of motion, and has been shown to apply to digital cursors as well as human movement:
$a + b\cdot \log _{2}{2D/W}$.

Given the bounding boxes of two widgets, we estimate $D$ as the distance between their centroids, and $W$ as the minimum of the target widget's box width and height~\cite{mackenzie1992extending}.  

\begin{figure}
    \centering
    \includegraphics[width=.8\columnwidth]{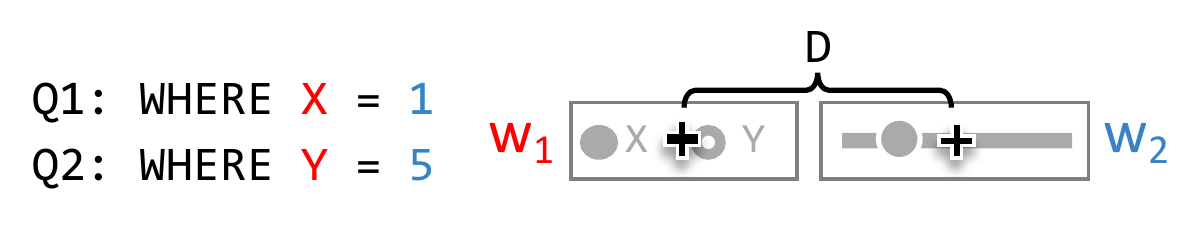}
    \vspace{-.15in}
    \caption{User navigates from $w_1$ to $w_2$ to change \texttt{Q1} to \texttt{Q2} . }
    \label{f:fitts}
\end{figure}

\begin{example}
  \Cref{f:fitts} shows two truncated queries that differ in the left and right operands of the equality predicate.  
  To express \texttt{Q1}, the user manipulates $w_1$ and $w_2$ to select \texttt{X} and \texttt{Y}, and again to express \texttt{Q2}, and navigates from $w_1\to w_2\to w_1\to w_2$.
  In the Fitts' law model, $D$ is the distance between the \texttt{+} markers, and $W$ is set of the box heights.
  Our prototype sets $a=1$ and $b=25$ based on manual experimentation.
\end{example}

We compute the navigation cost needed to express the input queries in sequence, and navigate the widgets in order of their depth first traversal in the \difftrees.

\stitle{Layout:}
The navigation cost implicitly takes the layout into account, as excessively wide or tall interfaces
will require more costly navigation.
Thus, by default we set the layout cost $C_L=0$.
However, if the user specifies a maximum desired width and height, then we add a penalization 
term $C_L(\mathbb{I}) = \alpha * (max(0,\mathbb{I}.w-width)+max(0,\mathbb{I}.h-height))$ 
if the interface size exceeds the desired maximum.

\section{Interface Generation}
\label{s:search}

As we saw in \Cref{s:overview}, the same set of queries can be expressed by many \difftree structures, each of which can be mapped to many possible interfaces.  Given the definitions in the previous sections, we can now present the interface generation problem: 

\begin{problem}[Interface Generation Problem] \label{p:problem}
  Given input query sequence $Q$ and interface cost model $C(\mathbb{I})$, return an interface mapping $\mathbb{I}^*=(\mathbb{V},\mathbb{M},\mathbb{L})$ such that $\mathbb{I}$ expresses $Q$ and minimizes $C(\mathbb{I})$.
\end{problem}

We solve this problem using a search-based approach, where we initialize
a set of \difftrees, and iteratively transform the \difftrees and map them to candidate interfaces.
This section describes the set of transformation rules that defines the search space,
and then describes the search procedure based on Monte Carlo Tree Search (MCTS)~\cite{Coulom2006EfficientSA}.

\begin{figure}
    \centering
    \includegraphics[width=.9\columnwidth]{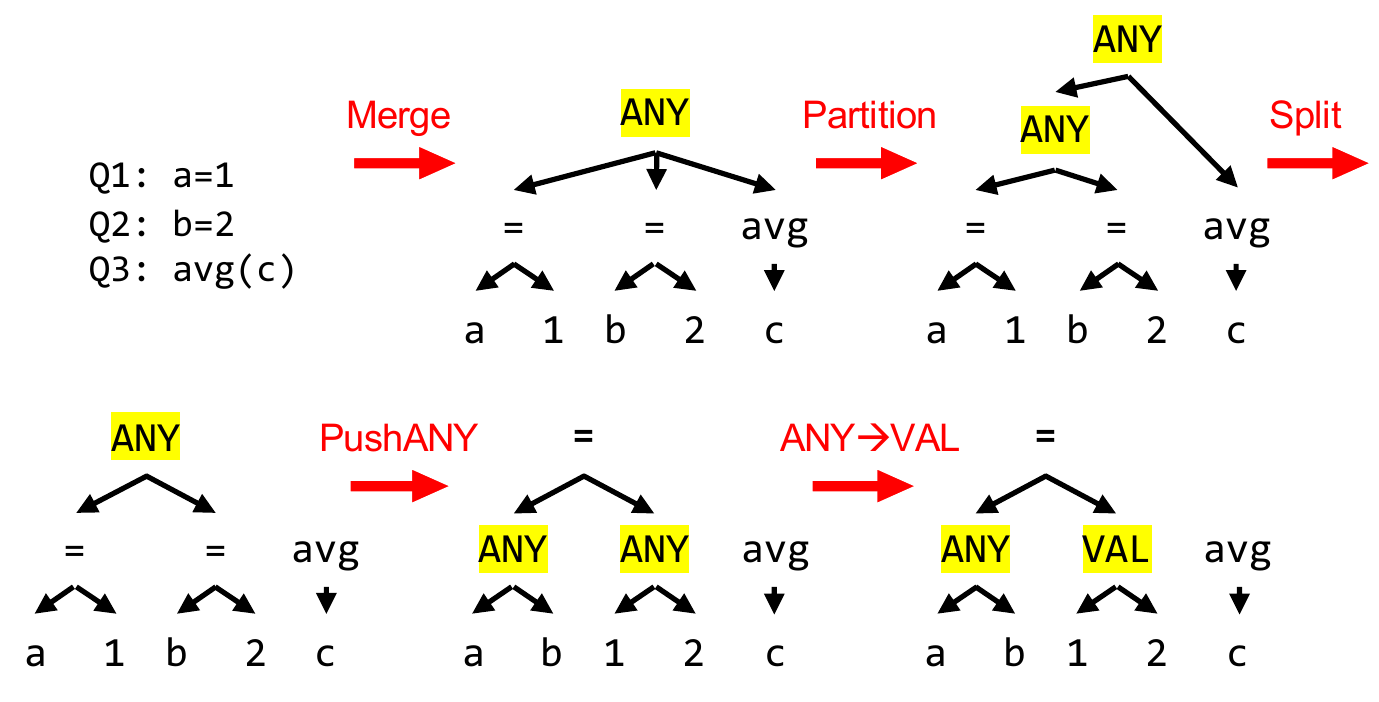}
	\caption{Example sequence of transformation rules applied to three input queries. }
    \label{f:transform}
\end{figure}

\subsection{\difftree Transformation Rules}
\label{ss:transformation}

We defined four categories of \difftree transformation rules that define the search space. 
Each rule takes as input a choice node and transforms the subtree rooted at the node.  
All rules are guaranteed to preserve or increase the expressiveness of the \difftrees;
since the initial set of \difftrees directly corresponds to the input queries, any reachable set of \difftrees
can also express those queries.

\begin{figure*}
  \centering
   \includegraphics[width=\textwidth]{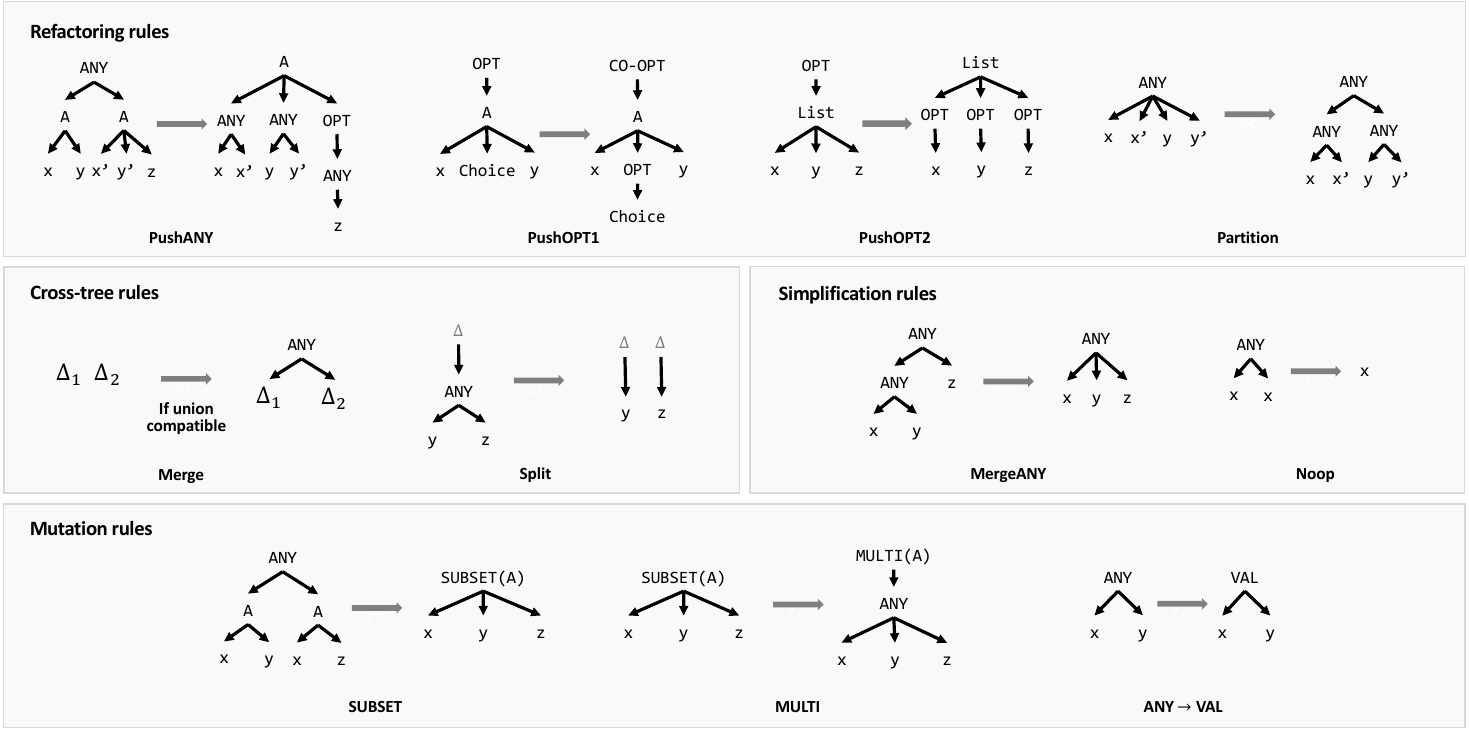}
 \caption{Four categories of \difftree transformation rules that define the search space.}
   \label{f:allrules}
\end{figure*}

\Cref{f:allrules} shows all the rules. In the diagram, \texttt
{x}, \texttt{y}, \texttt{z} represent subtrees that are
distinguished by their root node — the root of \texttt
{x} and \texttt
{x’} are
the same, and different than the \texttt{y}'s. \texttt{A} represents a AST node and  \textit{choice} represents \textit{Choice nodes}.

Each rule category serves a different purpose. 
{\it Refactoring rules} identify and refactor shared substructures in order to isolate the precise differences between the queries; these include \texttt{PushANY}, \texttt{PushOPT1}, \texttt{PushOPT2}, and  \texttt{Partition}.  
\texttt{PushANY} pushes \texttt{ANY} nodes down if their children have the same root node and introduces new \texttt{ANY} or \texttt{OPT} to express the differences between the children's children.
\texttt{Partition} which groups subsets of an \texttt{ANY} node's children, and \texttt{PushOPT1} and \texttt{PushOPT2}. \texttt{PushOPT1} pushes \texttt{OPT} node down to the \textit{Choice} node and leaves a new \texttt{CO-OPT} at the original place indicating that only if the \texttt{OPT} exists, the subtree rooted at \texttt{CO-OPT} exists. \texttt{PushOPT2} pushes the \texttt{OPT} down to all the children of the list node, which increases the expressiveness of the \difftrees. 
{\it Cross-tree rules} are used to \texttt{Merge} multiple \difftrees into one if their or \texttt{Split} one \difftree into multiple.
{\it Mutation rules} transform one type of choice node into another, for instance \texttt{ANY} $\rightarrow$ \texttt{VAL}, \texttt{MULTI}, and \texttt{SUBSET} in \Cref{f:allrules} .
Finally, {\it Simplification rules} are used to simplify the tree structure, for instance \texttt{Noop} removes \texttt{ANY} nodes with a single unique child, and \texttt{MergeANY} reduces a cascade of \texttt{ANY} nodes into a single one.

\begin{example}
  \Cref{f:transform} applies a sequence of transform rules on three input query fragments.
  Each query is initially a separate \difftree, and \texttt{Merge} combines them into a single \difftree.
  \texttt{Partition} groups the \texttt{ANY}'s children into homogenous clusters, and combines each non-singular cluster with an \texttt{ANY}.
  In practice, \texttt{Partition} is used to initially cluster the input queries by their result schema in order to reduce the number of
  redundant visualizations and maximize the likelihood of non-tabular visualization mappings.  
  \texttt{Split} then removes the root \texttt{ANY} so that its children are separate \difftrees.
  Since both equality predicates are rooted with \texttt{=}, \texttt{PushANY} pushes the \texttt{ANY}
  down.
  Finally, the \texttt{ANY} on the right has numeric children, so it is lifted to a \texttt{VAL}.
\end{example}

\subsection{Monte Carlo Tree Search}\label{ss:mcts}

A search problem is defined by its states, transitions, and cost function.  In our problem,
each set of \difftrees is a state, and transform rules define the transitions.
To assess the quality of each state, we apply the cost model in \Cref{ss:cost} to a set of $K$ random interface mappings.
The search returns a set of \difftrees, and we perform a more complete search for the final interface mapping $\mathbb{I}$.

Since the number of applicable transform rules is very large, it is infeasible to exhaustively search even a small portion of the search space.
Thus, we adopt Monte Carlo Tree Search~\cite{Coulom2006EfficientSA} (MCTS), a randomized search algorithm famously applied to problems with massive
search spaces, such as Google's AlphaGo~\cite{Silver2016MasteringTG}.
MCTS balances exploration of new states with exploitation of known good states. 
Another benefit is that MCTS works well when high-cost states are needed in order to reach an optimal state, as is common in games like Go. 
Since MCTS is traditionally used in two-player games, we use a single-player variation~\cite{schadd2008single}.

\subsubsection{MCTS Search Procedure}

Single-player MCTS~\cite{schadd2008single} explores the search space by iteratively growing a {\it search tree}, where each node is a search state\footnote{{\it State} refers to a search tree node in order to distinguish from a \difftree node.}.   Each iteration grows the tree using the following four steps.  Note that we add a special \texttt{TERMINATE} rule that is a valid transition for every state.  Choosing this rule results in a terminal state that has no outgoing transitions.
The algorithm stops when all leaves reach a terminal state.

\begin{enumerate}[leftmargin=*]
  
  \item \stitle{Select} a leaf state in the search tree.  Starting from the root state $R$, we recursively select a child until we reach a leaf state $S$.
Child selection is based on the Upper Confidence Bound for Trees strategy (UCT)~\cite{schadd2008single}: let $N_i$ be the $i^{th}$ child of current state $N$,
we choose the child that maximizes
{\small
\begin{align}
\overline{X}+c\sqrt{\frac{ln~t(N))}{t(N_i)}}+\sqrt{\frac{\sum{x^2}-t(N_i)\overline{X}^2+d}{t(N_i)}}
\label{eq:uct}
\end{align} }
Where $t(N)$ is the number of times that a node $N$ was visited.  This estimates an upper confidence bound for the expected reward $\bar{X}$, which is the negative cost.  The  term exploits the existing reward, the second term prefers the unexplored nodes, and the third term prefers nodes with high reward variance.  $c$ and $d$ are empirically set constants and control the preference for exploration and high variance nodes, respectively.

\item \stitle{Expand} the leaf $S$ by adding all of its children (result of valid transform rules) to the search tree (their visit count would be $0$, which will prioritize them in future iterations).  Then randomly choose a child state $C$ to simulate.

\item \stitle{Simulate} a random playout by applying random transform rules to $C$ until there are no more valid transform rules to apply, or if the \texttt{TERMINATE} rule is chosen.

\item \stitle{Backpropagate} the leaf node's reward from $C$ to $R$ by incrementing the visit count and appending the reward to all states in its path.  We estimate the reward by generating $K=5$ random interface mappings, estimating their costs, and returning the negative of the minimum cost (and thus the maximum reward).    
\end{enumerate}

\noindent In addition, we developed a set of optimizations that work well in practice.  The first is when choosing the best \difftree to return once the search terminates.  MCTS traditionally returns the state with the highest average reward.  In contrast, we follow Cadiaplayer~\cite{bjornsson2009cadiaplayer} and return the state with the maximum reward encountered (during its rollouts and random interface mappings).

The second is to run the search iterations in parallel.  Every $s$ iterations, the coordinator synchronously receives the highest reward state from each worker and distributes the maximum reward state back to the workers.    We further use early stopping, where each worker sends an early stop signal if its local optimal state has not changed in $es$ iterations.  If the coordinator receives stop signals from all workers and does not receive a higher reward state, it terminates the search.
\Cref{ss:perf} evaluates these optimizations.

\subsubsection{Interface Mapping Generation}
\label{ss:interfacemapping}

\begin{algorithm}
  \LinesNumbered
  \KwResult{ Given \difftrees $\Delta$ and queries Q, find the top k mappings of $\mathbb{V}$ and $\mathbb{M}$ with lowest $C_m$. }
  clist := an ordered list of choice nodes in $\Delta$\;
  icand := each choice node's all valid visualization interaction candidates\;
  wcand := each choice node's all valid widget candidates  \; 
  G[N] := the lowest $C_m$ among all widget covers of N\;
  F[N] := the top k exact widget covers of N with lowest $C_m$\;
  minHeap := the top k mappings of $\mathbb{V}$ and $\mathbb{M}$ with lowest $C_m$\;

  \SetKwFunction{FV}{searchV}
  \SetKwFunction{Fadd}{add}
  \SetKwFunction{Fdelete}{delete}
  \SetKwFunction{FM}{searchM}
  \SetKwFunction{FG}{G}
  \SetKwFunction{FF}{F}
  \SetKwProg{Fn}{Function}{:}{\KwRet}

  \Fn{\FG{N}}{
    \If{G[N] hasn't been calculated} {
      G[N] = min($C_m(w, Q)$ + \FG(N-w.cover) | w $\in$ wcand[[N[0]]])
    }
    \KwRet G[N]
  }
  \Fn{\FF{N}}{
    \If{F[N] hasn't been calculated} {
      L = $\emptyset$\;
      \For{w in wcand[N[0]]}{
        \If{w.cover $\subseteq$ N}
        {
          L = L $~\cup \{ C \cup \{w.cover \rightarrow w\} $ $| C \in \FF(N-w.cover)\}$
        }
      }
      F[N] = top K elements in $L$ with lowest costs
    }
    \KwRet F[N]
}

  \Fn{\FV{$\Delta$, $\mathbb{V}$, $\mathbb{M}$}}{
      \For{ v in all possible visualization mapping:}
      { 
        $\mathbb{V}$ = v\;
        compute icand\;
        searchM($\Delta$, 0, $\mathbb{V}$, $\mathbb{M}$ )\;
      }
  }

  \Fn{\FM{$\Delta$, $i$, $\mathbb{V}$, $\mathbb{M}$}}{
     N := the choice nodes without mapping in clist[0:i]\;
     Ns := all the choice nodes without mapping\;
     \SetNoFillComment 
     \tcc{pruning} 

    \If{$C_m(\{\mathbb{V}, \mathbb{M}\}, Q)$ + \FG(N) >= minHeap[k].cost}{\KwRet}
    \If{i == len(clist)}
    {
      \For {m in \FF(N)}
      {
        $\mathbb{M}$.\Fadd(m)\;
        \If{$C_m(\{\mathbb{V}, \mathbb{M}\}, Q)$ < minHeap[k].cost}
        {insert into minHeap\;}
        $\mathbb{M}$.\Fdelete(m)\;
     }
     \KwRet{}
    }
    \For{vinteraction in icand[i]}{
      \If{ vinteraction.cover $\subseteq$ Ns and compatible with $\mathbb{M}$}{
        $\mathbb{M}$.\Fadd(vinteraction)\;
      \FM($\Delta$, $i+1$, $\mathbb{V}$, $\mathbb{M}$)\;
      $\mathbb{M}$.\Fdelete(vinteraction)\;
      }
    }
    \FM($\Delta$, $i+1$, $\mathbb{V}$, $\mathbb{M}$)   \; 
  }
  \FV($\Delta$, $\emptyset$, $\emptyset$)\;
  \KwRet{minHeap}
   \caption{$\mathbb{V}$, $\mathbb{M}$ Mapping Generation Algorithm}
   \label{alg:interfacemapping}
  \end{algorithm}

Given the output state from MCTS, we perform a more exhaustive search for the lowest cost interface mapping.
This is done in three phases, where first, we enumerate all possible visualization mappings $\mathbb{V}$, and then derive all the valid visualization interactions over these visualizations. Afterwards, we search for all the mappings from the {\it choice nodes} to the widgets and visualization interactions $\mathbb{M}$. Finally, we construct the layout tree and use a prior branch-and-bound-based algorithm~\cite{Gajos2010AutomaticallyGP} to assign horizontal and vertical layouts to each layout node. Since manipulation cost $C_m$ term in the cost model is independent of the layout, and typically the dominant factor, we separate the search into two steps: first, we search for the top k mappings of $\mathbb{V},\mathbb{M}$ in terms of $C_m$, and second, we search the optimal layout for each of these top k mappings. In the end, we return the overall optimal interface.  We find $k = 10$ is sufficient to find the optimal interface empirically. 

\Cref{alg:interfacemapping} shows the pseudo code of our search algorithm for $\mathbb{V},\mathbb{M}$. At first, we enumerate $\mathbb{V}$ as shown in the \texttt{searchV} function(line 19). Then, we consider $\mathbb{M}$. 
An interaction mapping $\mathbb{M}$ is valid if and only if all the interactions exactly express the {\it choice nodes} once -- namely, to find the \textit{exact cover} of the {\it choice nodes} using widgets or visualization interactions.  Notice that, the attribute \texttt{cover}(line 15) of a choice node's candidate interaction means besides this choice node, all the choice nodes it expresses at the same time. For example, in \Cref{f:twochoice}, both \texttt{ANY} have \texttt{RangeSlider} as its candidate widget and \texttt{RangeSlider.cover} are these two \texttt{ANY}. 
 Also, visualization interaction mapping is more sophisticated in that \textcircled{1} one visualization interaction can not be mapped to multiple times in the same \difftree because it will decrease the expressiveness; \textcircled{2} on one visualization, some interactions are conflicted, such as brush along x-axis and brush along y-axis, so that only one of them can be chosen.  With such concerns, the interaction mapping can not be solved by single dynamic programming. Thus, we separate visualization interaction mapping and widget mapping -- first,  enumerate the compatible visualization interactions (line 36); and for each visualization interaction mapping, find the optimal widget mappings(line 30) for the left uncovered choice nodes using dynamic programming in \texttt{F(N)}(line 11). 
 Also, we prune the search space by proposing a lowest bound(line 27), which equals to the existing visualization interaction mapping's cost plus the lowest possible widget mapping cost for
 uncovered choice nodes computed by \texttt{G(N)}(line 7). 
When this lowest bound is greater than the existing $k_{th}$ optimal cost, we prune this branch.

\section{Experiments}\label{s:exp}

The primary success criteria for \sys is to generate fully functional
interactive visualizations from a small number of input examples.  
We break this down into three questions.
1) Can \sys generate expressive interactive visualization interfaces?  To evaluate this,
we follow the evaluation in Vega-lite~\cite{Satyanarayan2017VegaLiteAG} and show examples that cover the data-oriented
interactions in Yi et al.'s~\cite{Yi2007TowardAD} taxonomy of interaction methods.  
We further show that \sys can reproduce the COVID visualization shown at the top of Google's search results page 
for the search query ``covid19''.
2) How \sys help interface creation in realistic settings?  We illustrate this by using a subset of queries from the Sloan Digital Sky Survey~\cite{york2000sloan} (SDSS) to produce a custom interface.  We also show a case study to create an analysis dashboard for the Kaggle supermarket sales dataset~\cite{kagglesales} from complex sales analysis queries that existing authoring tools (e.g., Metabase, Tableau) do not support.    
3) How quickly can \sys generate interfaces?  We evaluate \sys's latency and generated interface quality varies with respect to its search parameters.  

Our focus is on the {\it functionality} of the output interface (whether it can easily express the underlying analysis) rather than its style and presentation.   Thus, when applicable, we may modify the font, spacing, and other CSS-based interface stylings.

\subsection{Interaction Expressiveness}\label{e:interactionexp}
We use Yi et al.`s~\cite{Yi2007TowardAD} taxonomy of visualization interaction techniques to highlight \sys's expressiveness.
Their paper describes seven interaction types:
{\it Select} interesting data;
{\it Explore:} show different subsets of the data;
{\it Abstract:} change the level of detail;
{\it Filter} the data;
{\it Connect:} highlight related data (such as in a different chart);
{\it Encode:} change the visual representation (e.g., from scatterplot to bar chart);
{\it Reconfigure:} rearrange the visual presentation (e.g., change from linear to log scale)
Of these, encode and reconfigure are unrelated to query-level transformations.
Every example supports selection, so we evaluate the remaining four types.  

In each example, we \highlight{highlight} query fragments that differ from the 
preceding query, use \texttt{..} when a long substring does not change, and use \texttt{BTWN min \& max} to mean \texttt{BETWEEN min AND max}.

\begin{figure*}
  \begin{tabular}[t]{ccc}
    \begin{tabular}{c}
    \begin{subfigure}[t]{.32\textwidth}
      \includegraphics[width=.9\linewidth]{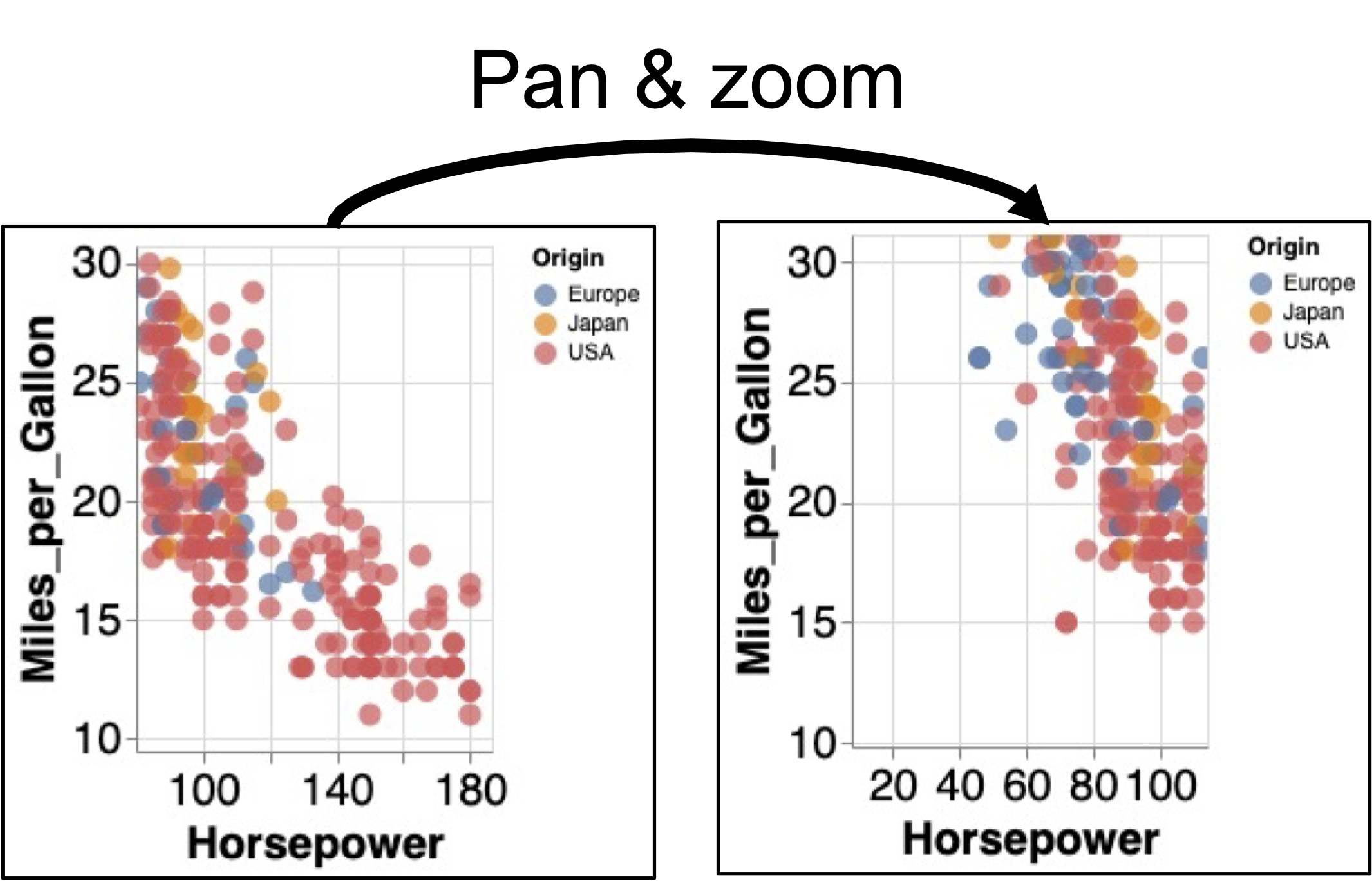}
      \caption{Explore}
      \label{f:explore}
    \end{subfigure}\\
    \begin{subfigure}[t]{.32\textwidth}
        \includegraphics[width=\linewidth]{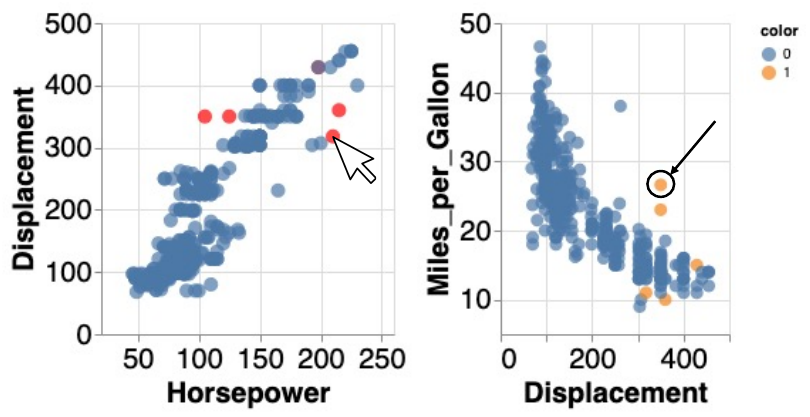}
        \caption{Connect}
        \label{f:connect}
    \end{subfigure}
    \end{tabular}
    &
    \begin{subfigure}{.32\textwidth}
      \includegraphics[width=.9\linewidth]{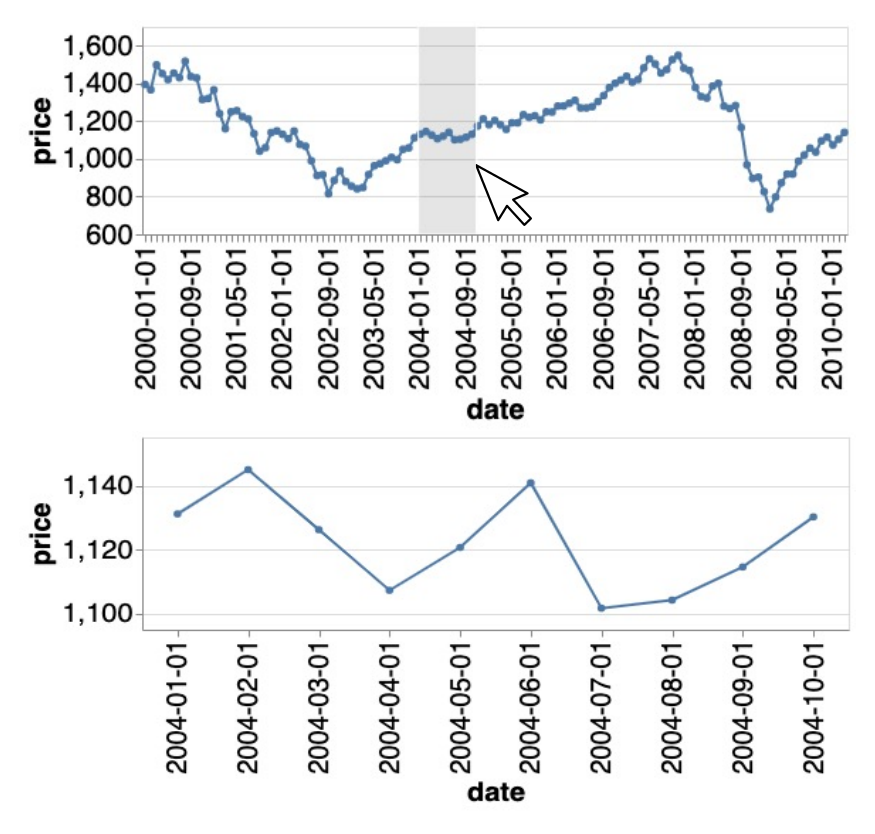}
      \caption{Abstract}
      \label{f:abstract}
    \end{subfigure}
    &
    \begin{subfigure}{.30\textwidth}
      \includegraphics[width=.9\linewidth]{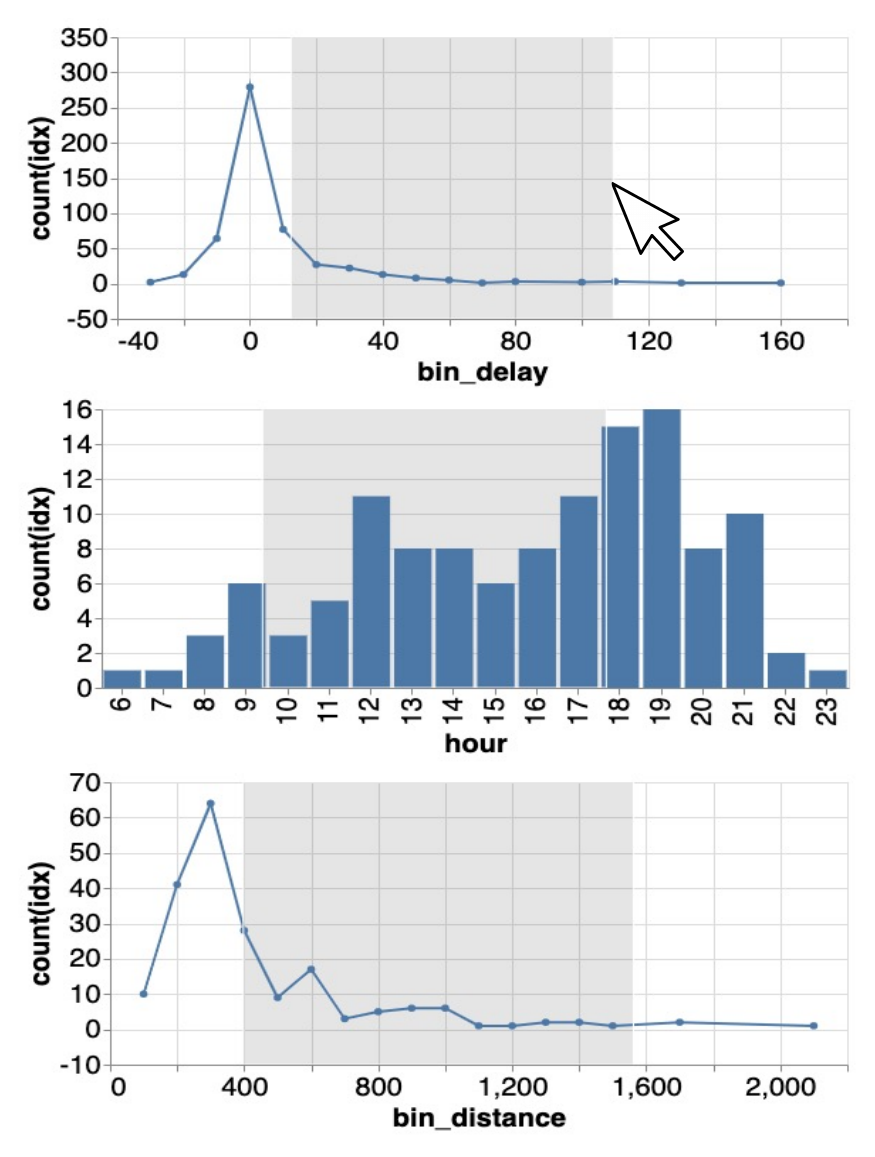}
      \caption{Filter}
      \label{f:filter}
    \end{subfigure}
  \end{tabular}
  \vspace{-.1in}
  \caption{Interfaces that express Yi et al.'s~\protect\citeA{Yi2007TowardAD} interaction taxonomy using queries in \Cref{l:explore}, \Cref{l:connect}, \Cref{l:overview} and \Cref{l:filter}.  (a) panning ans zooming interaction, (b) linked selection, (c) overview and details interaction, (d) cross-filtering.}
\end{figure*}

\begin{figure*}
  \centering
  \begin{tabular}[t]{ccc}
    \begin{subfigure}[b]{.30\textwidth}
      \centering
      \includegraphics[width=\linewidth]{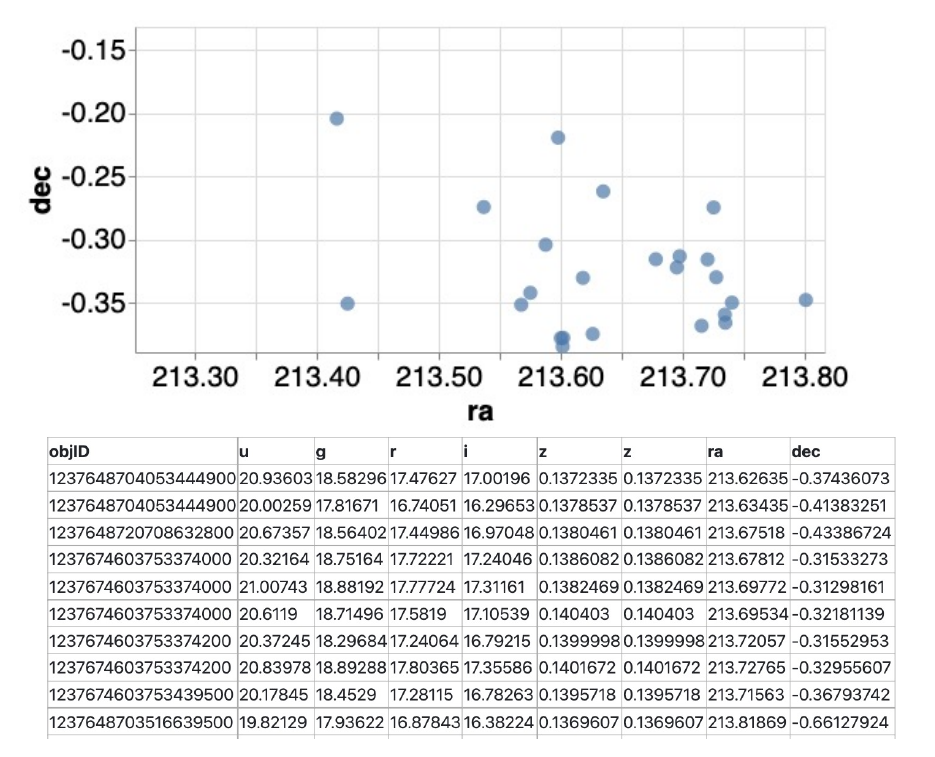}
      \caption{}
      \label{f:sdss}
    \end{subfigure}
    &
    \begin{subfigure}[b]{.33\textwidth}
      \centering
      \includegraphics[width=\linewidth]{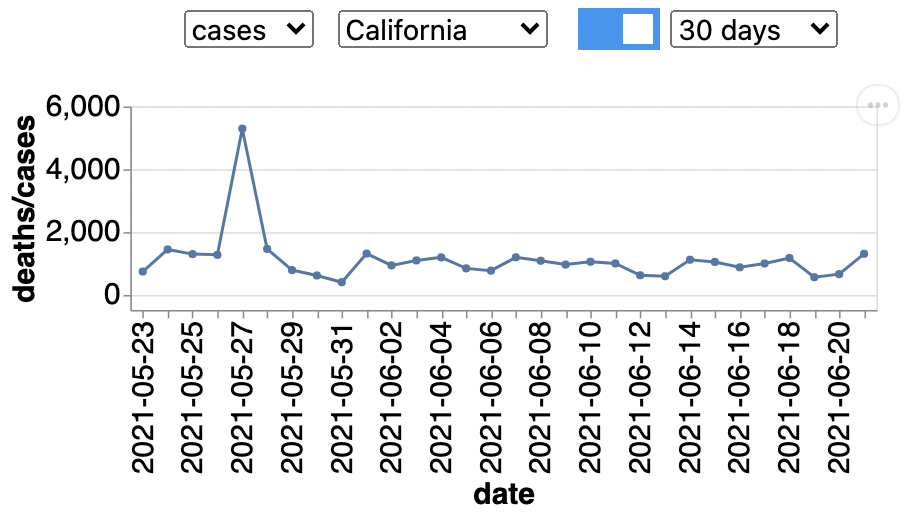}
      \caption{}
      \label{f:covid}
    \end{subfigure}
    &
    \begin{subfigure}[b]{.30\textwidth}
      \centering
      \includegraphics[width=.8\linewidth]{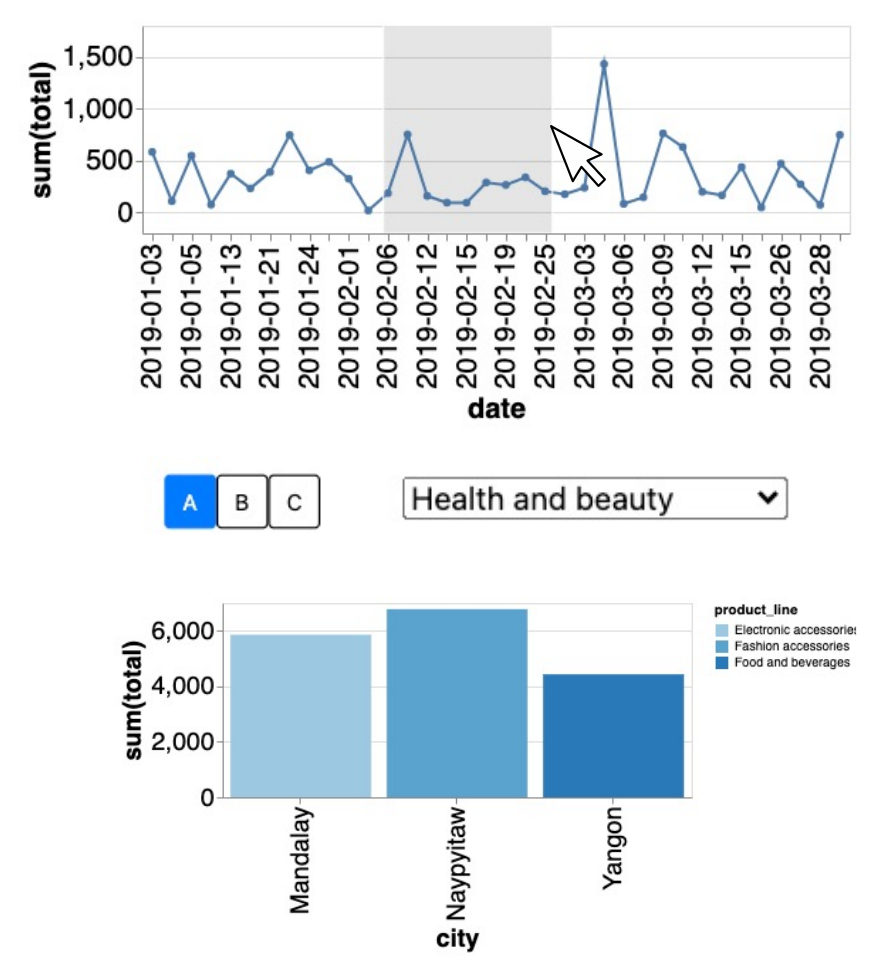}
      \caption{}
      \label{f:sales}
    \end{subfigure}
  \end{tabular}
  \vspace{-.1in}
  \caption{Interfaces generated for case studies. (a) new interface from real SDSS queries, (b) reproducing Google's Covid-19 Vis, (c) authoring a custom sales analysis dashboard.  }
\end{figure*}

  \begin{lstlisting}[
  caption = { Explore}, 
  label={l:explore}]
Q1  SELECT hp, mpg, origin from Cars
     WHERE hp BTWN 50 & 60 AND mpg BTWN 27 & 38
Q2 ..WHERE hp BTWN ^\highlight{60}^ & ^\highlight{90}^ and mpg BTWN ^\highlight{16}^ & ^\highlight{30}^
  \end{lstlisting}

\stitle{Explore: }  The queries in \Cref{l:explore} project horsepower (hp), miles per gallon (mpg), and origin from the \texttt{Cars} dataset, and change the range predicates on hp and mpg.  The generated interface in \Cref{f:explore} renders the attributes as the x-axis, y-axis, and color, respectively, and enables panning and zooming to control the range predicates (thus it also satisfies {\it Abstract} interaction described next).  

  \begin{lstlisting}[
  caption = {Abstract}, 
  label={l:overview}]
Q1 SELECT date, price FROM sp500
Q2 ..^\highlight{WHERE date > '2001-01-01' AND date < '2003-01-01'}^
Q3 ..WHERE date > ^\highlight{'2001-02-01'}^ AND date < ^\highlight{'2003-02-01'}^
  \end{lstlisting}

\stitle{Abstract } The queries in \Cref{l:overview} also vary a range predicate, however \texttt{Q1} does not have a \texttt{WHERE} clause.  The overview-and-detail interface (\Cref{f:abstract}) has a static overview chart for \texttt{Q1}; brushing it updates the detail line chart that expresses the filtered data.

  \begin{lstlisting}[
  caption = { Connect}, 
  label={l:connect} ]
Q1 SELECT hp, disp, id FROM Cars
Q2 SELECT mpg, disp, ^\highlight{id in (1, 2) as color}^ FROM Cars
Q3 SELECT mpg, disp, id in ^\highlight{(20,22)}^ as color FROM Cars
  \end{lstlisting}

\stitle{Connect:} \texttt{Q1} in \Cref{l:connect} returns horsepower (hp), displacement (disp), and id from the \texttt{Cars} dataset, while the latter queries return miles per gallon (mpg), and a boolean color attribute based on the ids of the rows.   In \Cref{f:connect}, the left scatterplot renders hp and disp as the x and y axes (id is a primary key so is not rendered by default), while the right scatterplot also encodes the boolean color attribute as the mark color.   Thus, selecting points in the hp chart highlights the corresponding rows in the mpg chart.
  \begin{lstlisting}[
  caption = {Filter}, 
  label={l:filter}, breaklines=false]
Q1 SELECT hour,count(*) FROM flights GROUP BY hour
Q2 ..^\highlight{WHERE delay BTWN 0 \& 50 AND dist BTWN 400 \& 800}^..
Q3 ..WHERE delay BTWN ^\highlight{10}^ & ^\highlight{60}^ AND dist BTWN ^\highlight{10}^ & ^\highlight{300}^..
 
Q4 SELECT ^\highlight{delay}^,count(*) FROM flights GROUP BY ^\highlight{delay}^
Q5 ..^\highlight{WHERE hour BTWN 10 \& 16 AND dist BTWN 400 \& 800}^..
Q6 ..WHERE hour BTWN ^\highlight{15}^ & ^\highlight{20}^ AND dist BTWN ^\highlight{200}^ & ^\highlight{700}^..
 
Q7 SELECT ^\highlight{dist}^, count(*) FROM flights group by ^\highlight{dist}^
Q8 ..^\highlight{WHERE hour BTWN 10 \& 16 AND delay BTWN 0 \& 50}^..
Q9 ..WHERE hour BTWN ^\highlight{8}^ & ^\highlight{19}^ AND delay BTWN ^\highlight{20}^ & ^\highlight{61}^..
  \end{lstlisting}

\stitle{Filter:}  \Cref{l:filter} lists three sets of queries; each is grouped by a different attribute (\texttt{hour}, \texttt{delay}, and \texttt{dist}).   \texttt{Q1,4,7} do not filter the table, and the subsequent queries filter by the grouping attributes of the other two sets.  For instance, \texttt{Q2,3} group on \texttt{hour} and filter on \texttt{delay} and \texttt{dist}.  These queries describe  cross-filtering, which \sys automatically derives from first principles (\Cref{f:filter}).  Brushing in a chart updates the corresponding predicates in {\it both} other charts, and clearing the brush disables the predicate (toggles its existence).

\subsection{Case Studies}

We show case studies that use a real-world query log, reproduce a real-world visualization, and author a complex sales dashboard.

\begin{lstlisting}[caption = {Subset of SDSS queries}, label={l:sdss}]
Q1 SELECT DISTINCT gal.objID, gal.u, gal.g, gal.r, 
          gal.i, gal.z, s.z, s.ra, s.dec 
   FROM galaxy as gal, specObj as s 
   WHERE s.bestObjID = gal.objID AND s.z BTWN 0.1362 & 0.141 AND 
         s.ra BTWN 213.3 & 214.1 AND s.dec BTWN -0.9 & -0.2
Q2 ..AND s.ra BTWN ^\highlight{213.4191}^ & ^\highlight{213.9}^ AND s.dec BTWN ^\highlight{-0.565}^ & ^\highlight{-0.3111}^
Q3 ..AND s.ra BTWN ^\highlight{213.5}^ & ^\highlight{213.8}^ AND s.dec BTWN ^\highlight{-0.34}^ & ^\highlight{-0.2}^
        -- many similar queries --
Q8 ^\highlight{select DISTINCT ra, dec FROM specObj}^
     ^\highlight{WHERE ra BTWN 213.2 \& 213.6 AND dec BTWN -0.3 \& -0.1}^
Q9 ..WHERE ra BTWN ^\highlight{213}^ & ^\highlight{214}^ AND dec BTWN ^\highlight{-0.8}^ & ^\highlight{-0.4}^
\end{lstlisting}

\stitle{SDSS queries: }
Visitors can use textbox-based forms on the SDSS website~\cite{sdssradial} to select a subset of stars that are returned as a text table.  The non-interactive forms are complex to support a wide range of analyses.   \sys uses a subset of SDSS queries~\cite{sdsslogused} (\Cref{l:sdss}) to generate a custom analysis interface.
\texttt{Q1} is a join query to filter stars by their celestial coordinates (z, ra, dec) and \texttt{Q2-7} vary the right ascension (ra) and declination (dec).  Finally, \texttt{Q8,9} return star locations within a bounding box.
\Cref{f:sdss} renders the first set of queries as a table because those queries return 9 attributes, and renders the star locations (\texttt{Q8,9}) as a scatterplot.  Users can pan and zoom in the scatterplot to update the table.  
In short, \sys transformed a text-based form into a fully interactive visual analysis interface.

\stitle{Reproducing Google's Covid-19 Vis} 
This example uses queries in \Cref{l:covid} to reproduce the Covid-19 visualization on Google's results page for ``covid19''.
\texttt{Q1-3} compute daily confirmed cases for different states and date intervals, while \texttt{Q4-8} report daily deaths.  Note that \texttt{Q1,Q4} do not filter by date.
\Cref{f:covid} reproduces the interactions in Google's visualization. The dropdowns change the reported metric, state filter, and date interval.  The latter is an example of a nested interaction, because the filter on date interval dropdown is only enabled when the toggle is turned on.
\begin{lstlisting}[
  caption = {Covid visualization queries}, label={l:covid}, breaklines=false]
Q1 SELECT date, cases FROM covid WHERE state='CA'
Q2 ..WHERE state=^\highlight{'WA' and date>date(today(), '-30 days')}^
Q3 ..WHERE state=^\highlight{'CA'}^ and date>date(today(), ^\highlight{'-7 days'}^)
Q4 SELECT date, ^\highlight{deaths}^ FROM covid WHERE state='CA'  ^\highlight{\hspace{2em}}^
Q5 ..WHERE state=^\highlight{'NY'}^
Q6 ..WHERE state='WA' ^\highlight{and date>date(today(),'-14 days')}^
Q7 ..WHERE state='WA' and date>date(today(),^\highlight{'-7 days'}^)
Q8 ..WHERE state=^\highlight{'NY'}^ and date>date(today(),'-7 days')
\end{lstlisting}

\begin{lstlisting}[caption = {Complex sales analysis queries}, label={l:sales}]
Q1   SELECT city, product, sum(total) FROM sales as ss 
     WHERE ss.date 
     GROUP BY city, product 
     HAVING sum(total) >= ( SELECT max(t) FROM 
       (SELECT sum(total) as t FROM sales as s 
        WHERE s.city = ss.city and  
        GROUP BY s.city, s.product ) )
Q2 ..WHERE ss.date ^\highlight{BTWN '2019-01-25'}^ & ^\highlight{'2019-02-15'}^
..HAVING sum(total) >= ( SELECT max(t)
            ..s.date ^\highlight{BTWN '2019-01-25'}^ & ^\highlight{'2019-02-15'}^ ..'
Q3 ..WHERE ss.date BTWN ^\highlight{'2019-01-25'}^ & ^\highlight{'2019-02-15'}^
   ..HAVING sum(total) >= ( SELECT max(t)
            ..s.date BTWN ^\highlight{'2019-01-25'}^ & ^\highlight{'2019-02-15'}^ ..
Q4   ^\highlight{SELECT date, sum(total) FROM sales }^
     ^\highlight{WHERE branch = 'A' AND product = 'Health and beauty'}^
     ^\highlight{GROUP BY date}^
Q5   ..WHERE branch = ^\highlight{'B'}^ and product = ^\highlight{'Electronics'}^..
Q6   ..WHERE branch = ^\highlight{'C'}^ and product = ^\highlight{'Lifestyle'}^..
        -- many similar queries --
\end{lstlisting}

\stitle{Authoring a Sales Dashboard:}
This example is an analysis that current authoring tools can't create.  Query 1 in \Cref{l:sales} returns the total sales for products in different cities with the maximum total sales; it has multiple nested queries in the \texttt{HAVING} clause.  \texttt{Q2} modifies the query by specifying the date range, \texttt{Q3} modifies \texttt{Q2}'s date predicate in the outer and nested queries, and the remaining return total sales by date for different branches and products.  
The top chart in \Cref{f:sales}  renders the total sales by date, with controls to filter by branch and product; brushing it updates the bar chart (which renders the product with top sales for each city during the time period specified by the brush).  \sys can transform arbitrarily complex queries, and link visualizations.   Existing authoring tools cannot generate this interface: Metabase only supports parameters in the \texttt{WHERE} clause~\cite{metabasedoc}, and Tableau does not parameterize custom queries~\cite{tableaudoc}.

\subsection{\sys Performance and Quality}\label{ss:perf}

\sys is primarily affected by three parameters:

\textbf{Early Stop: } stop MCTS if the optimal \difftree doesn't change in $es\in[5,100]$ iterations (default 30).

\textbf{Parallelize} over $p\in[1,4]$ workers (default 3).

\textbf{Synchronization Interval: } every $s\in[5,100]$ MCTS iterations (default 10).

We measure runtimes for MCTS search and the final interface mapping separately.  We report the interface quality as follows: given $c$ as an interface's cost, and $c^*$ as the minimum cost over all evaluated conditions for a query log, we report $\frac{c^*}{c}$.   $1$ means the generated interface is optimal, and worse interfaces converge towards $0$.  We visually verified that interfaces with $c^*$ were indeed the best, and qualitatively, interfaces with quality above $85\%$ are nearly the same as the optimal (see appendix for examples of interfaces of varying qualities).  

We used all 7 query logs above, and average over 10 runs/condition on 4x2.2GHz 16GB RAM Google Cloud VMs with Ubuntu 20.04 LTS.

\stitle{Runtime-Quality Trade-off: }
We first report the end-to-end runtime and interface quality from a sweep of all the parameters.  We vary $es$ and $s$ between 5 to 100 in increments of 5, and vary the parallelization from 1 to 4.  We find that \sys is able to find the optimal interface in less than $1$ second for the ``simpler'' query logs (e.g., Explore, Connect, Sales, SDSS).  Filter and Covid are more challenging because they result in many visualizations or interactions, and we see a general trade-off where configurations with longer runtimes tend to generate higher quality results.   We will dive into the sensitivity to each parameter below.  To avoid crowding the graphs, we will use Explore to represent the ``simpler'' logs.

\begin{figure}
 \centering
 \includegraphics[width=\columnwidth]{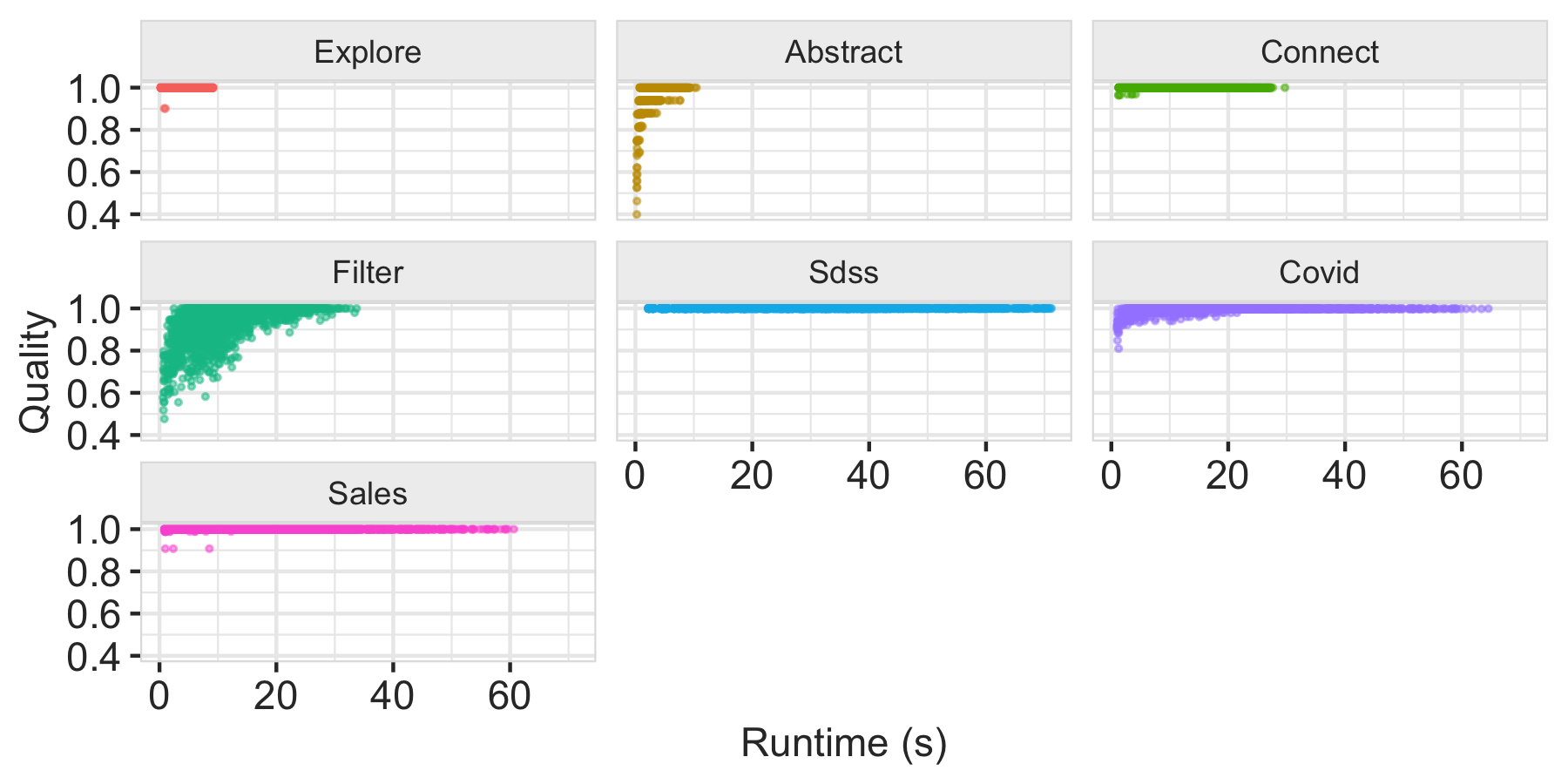}
 \caption{Runtime-quality trade-offs across all conditions.} 
 \label{f:quality-time}
\end{figure}

\begin{figure}
	\centering
	\includegraphics[width=\columnwidth]{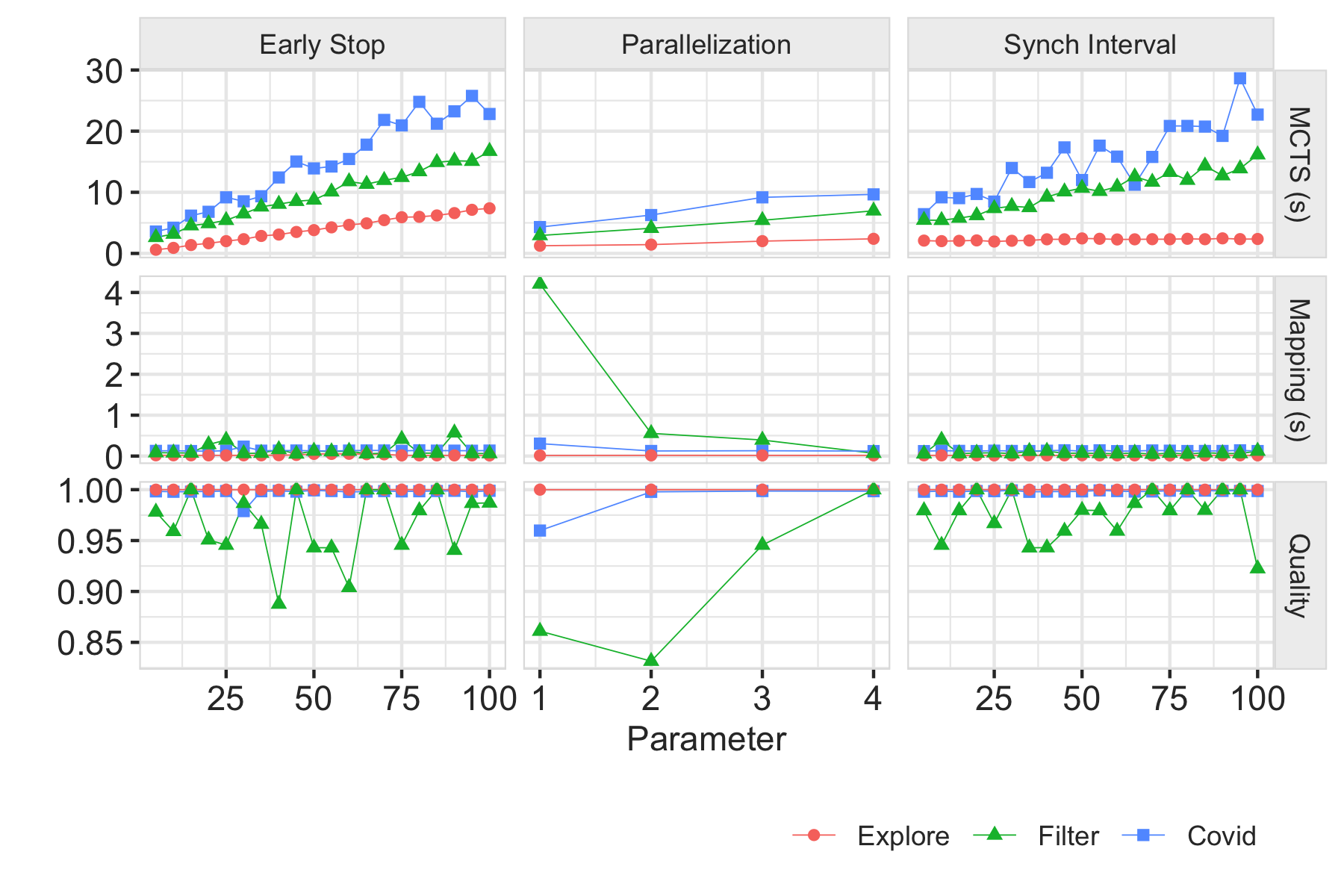}
    \vspace{-.2in}
	\caption{Varying early stop and synchronization interval increases runtime without tangible improvements to interface quality.  In contrast, parallelization increases the MCTS cost, but can find higher quality \difftree structures for complex interfaces such as Filter.}
	\label{f:all}
  \end{figure}

\stitle{Parameter Sensitivity}
\Cref{f:all} reports the three metrics (rows) while varying each parameter (cols).
To keep the plots legible, we report results for Explore, Filter, and Covid because the remaining logs have results nearly identical to Explore. 
Varying early stop and the synchronization interval increases the MCTS runtime, but does not impact interface quality.  This is because \sys finds the optimal \difftree structure very quickly, so larger early stop values or synchronization intervals simply delay MCTS termination.  

Varying parallelization (middle col) slows MCTS search due to synchronization overhead and stragglers.  The mapping cost for Filter when there is no parallelization is higher because the poor quality \difftree contains a huge number of choice nodes that must be mapped.  High quality \difftrees tend to have fewer choice nodes, which reduces the number of redundant interactions.   Increasing parallelization improves quality for Filter because MCTS explores a larger subset of the search space.  Note that the y-axis for quality ranges from $85\%$ to $100\%$, thus even with low parallelization, the interfaces have reasonable quality.

\stitle{Scalability:}
We also evaluated the runtime as the number of input queries increases from 9 to 900 (by duplicating the Filter log).  We find that the runtime increases roughly linearly from a few seconds to ${\approx}2000$s for 900 queries.   This is expected as \sys is not optimized for scalability, and is dominated by (1) increased number of search states, (2) higher cost to estimate the navigation cost due to the larger number of queries, and (3) increased cost to check safety.   We expect sampling, caching, and approximation optimizations can reduce these bottlenecks considerably.  Further, we note that in an authoring setting, the number of queries is unlikely to be large.

\section{Related Work}\label{s:related}

\stitle{Interface generation:}
Existing works either take analysis queries into account, nor generate fully interactive analysis interfaces.
For instance, prior work in the DB community generates form-based search and record creation interfaces based solely on the database content~\cite{jayapandian2008automated,Jagadish2007MakingDS,Jayapandian2006AutomatingTD}, but may generate over-complex forms because it does not leverage analysis queries.
Similarly, techniques from the HCI community rely on the developer to provide task and data specifications~\cite{puerta1994model,vanderdonckt1994automatic,nichols2004improving,swearngin2020scout,Gajos2004SUPPLEAG}. 
In this sense, \sys models the desired task using example queries.
Visualization recommendation algorithms~\cite{wang2021falx,Zong2019Lyra2D,Moritz2018FormalizingVD,wongsuphasawat2016voyager,mackinlay2007show,mackinlay1986automating} output visualization designs based on an input dataset but not full analysis interfaces.
\sysold~\cite{zhang2019mining} uses input queries, but is limited to unordered sets of widgets.

\stitle{Authoring Tools:}
There are numerous interface~\cite{jquery,react,chang2015shiny} and visualization~\cite{Satyanarayan2017VegaLiteAG,Bostock2011DDD} programming libraries available, and 
tools like AirTable~\cite{airtable}, Figma~\cite{figma}, and Plasmic~\cite{plasmic} help interface designers rapidly iterate on the interface design.
However, these still require programming effort to use and combine.
Dashboard authoring tools such as Metabase~\cite{metabase}, Retool~\cite{retool}, and Tableau~\cite{Murray2013TableauYD} let non-technical users create interactive SQL-based analysis dashboards.
However, they restrict types of analyses and queries (e.g., to parameterized literals or OLAP queries) in order to keep their own interfaces simple.  
In contrast, \sys supports arbitrarily complex queries and structure transforms for nearly any syntactic element in the queries.  Users ``program'' \sys by providing example queries.

\section{Discussion and Conclusion}
\label{s:conclusion}

\sys is the first system to generate fully functional interactive visualization interfaces from a few examples analysis queries.  \sys introduces the \difftree structure to succinctly encode syntactic variations between input queries as {\it choice nodes}.  It formulates interface generation as a schema matching problem from \difftree results to visualizations, choice nodes to interactions, and \difftree structure to layout.  Transform rules ``refactor'' the \difftrees to produce different interfaces.
\sys can generate interfaces to cover the interaction taxonomy proposed by Yi et al.~\cite{Yi2007TowardAD}, can generate useful interfaces from real-world query logs, and replicate existing visualizations, and can be used to author visualizations that are not possible in existing visual authoring tools.  In the evaluation, \sys generated interfaces in $2-19$s, with a median of 6s.  

Our future work plans to improve system usability so users can control how much \sys generalizes from the input queries, and replace subsets of the interface they do not like.  Further, we will focus on generating informative labels, support design principles such as alignment and spacing, and improve scalability.

\balance

\bibliographystyle{ACM-Reference-Format}
\bibliography{main}  %

\appendix
\addappheadtotoc
\section{examples of interfaces of varying qualities}

\begin{figure}[H]
    \centering
    \includegraphics[width=0.55\columnwidth]{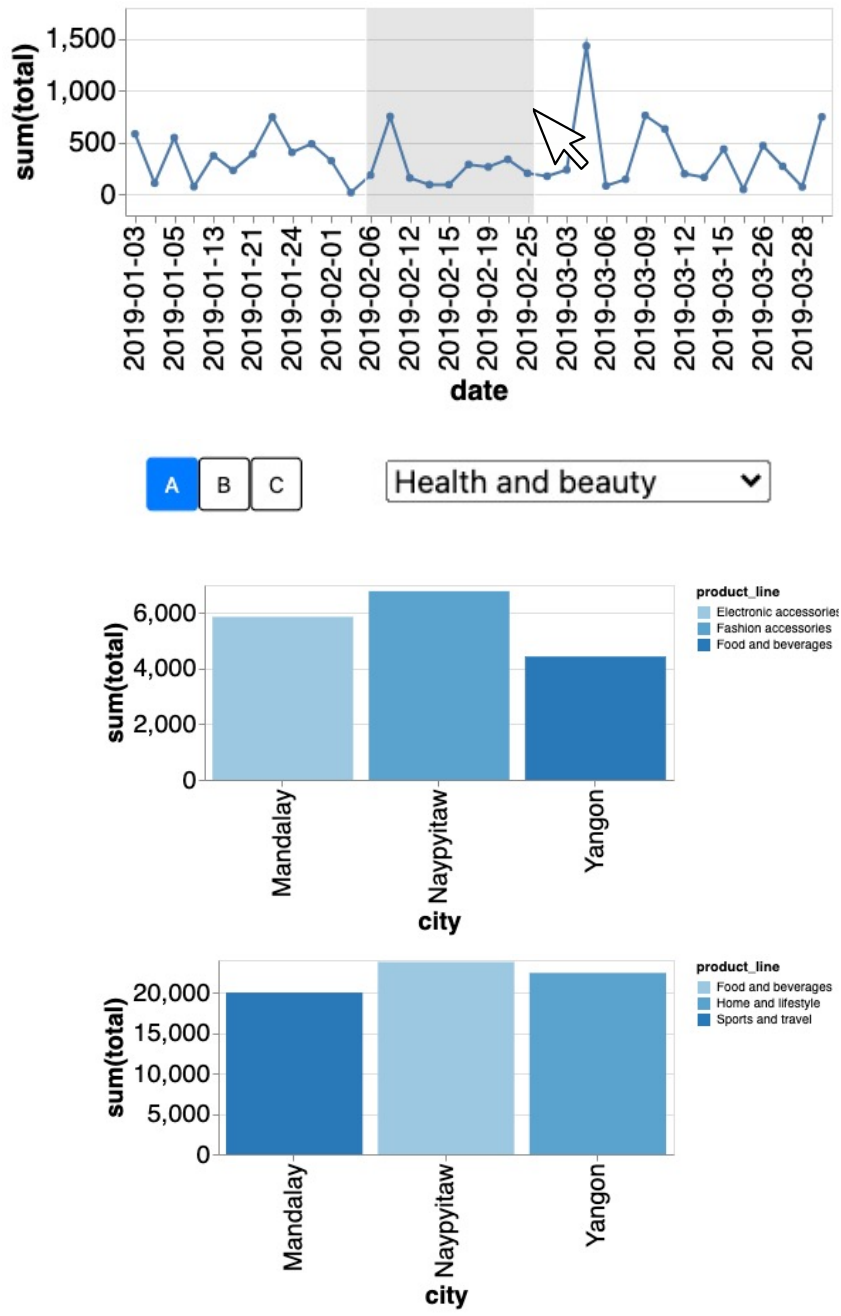}
    \caption{Non-optimal interface for sales analysis queries with quality = 0.893.} 
    \label{f:sales-0.893}
\end{figure}

Below shows two non-optimal interfaces. \Cref{f:filter-0.87} shows an interface
for filter queries in \Cref{l:filter} with its quality equal to 0.87. Compared with the optimal
interface in \Cref{f:filter}, it has an extra toggle button which toggles
the existence of the first chart's filter predicates. If the
toggle  is on, the brushing in the second and third charts will update the
first chart by adding corresponding predicates to its underlying query's where clause,
otherwise, the first chart will not be affected by the brushing
interactions on the other charts. \Cref{f:sales-0.893} shows the interface for
sales analysis queries in \Cref{l:sales} whose quality is 0.893. Compared to the optimal one in
\Cref{f:sales}, the third static chart is newly added which only express
\texttt{Q1} -- the total sales for products in different cities with the maximum
total sales \textit{without} any date range constraint. The other charts stay the same that
brushing in the first chart will update the second chart by specifying the date
range. From above, we can see that although we did not find the optimal \difftrees in these two example, such non-optimal interfaces with high quality above $85\%$ are already nearly the same as the optimal ones . 

\begin{figure}[H]
    \centering
    \includegraphics[width=0.55\columnwidth]{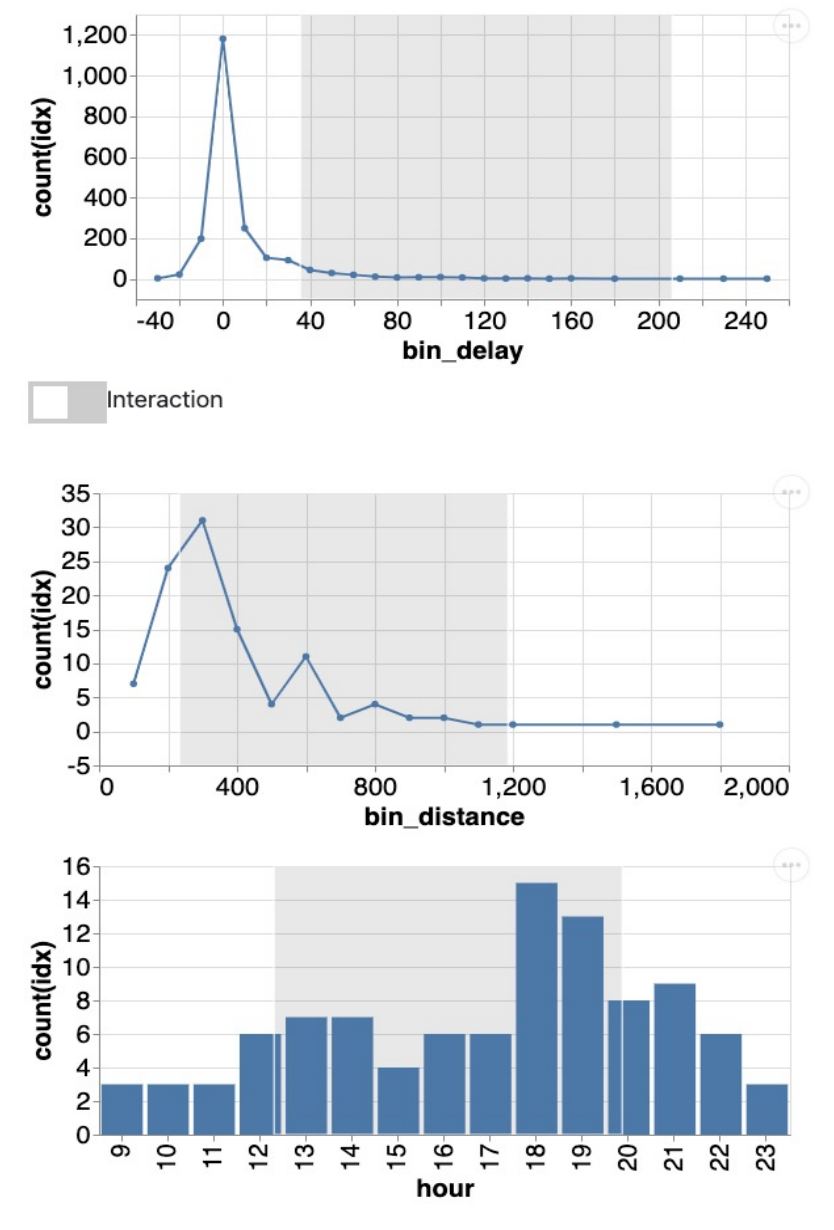}
    \caption{Non-optimal interface for filter usecase with quality = 0.87. } 
    \label{f:filter-0.87}
\end{figure}

\end{document}